\documentclass[reprint, aps, amsmath, amssymb, pra, 10pt, floatfix, notitlepage]{revtex4-1}

\usepackage[english]{babel}   
\usepackage[pdftex]{graphicx}
\usepackage{amsmath, color, colordvi}
\usepackage{dcolumn}
\usepackage{bm}
\definecolor{navyblue}{rgb}{0.0, 0.0, 0.5}

\usepackage{hyperref}
\usepackage[draft]{changes}
\hypersetup{
    bookmarks=false,         
    unicode=false,          
    pdftoolbar=false,        
    pdfmenubar=true,        
    pdffitwindow=false,     
    pdfstartview={FitH},    
    pdftitle={},    
    pdfauthor={},     
    pdfsubject={},   
    pdfcreator={},   
    pdfproducer={}, 
    pdfkeywords={many-body localization} {non-ergodic dynamics}{bosons}{level statistics}{thermalization}, 
    pdfnewwindow=true,      
    colorlinks=true,       
    linkcolor=black,          
    citecolor=blue,        
    filecolor=magenta,      
    urlcolor=blue           
}

\usepackage[protrusion=true, expansion=true]{microtype} 
\usepackage{braket}
\usepackage[per-mode=symbol, separate-uncertainty, range-phrase=--, range-units=single]{siunitx}
\usepackage{multirow}

\DeclareMathOperator{\Tr}{Tr}

\usepackage{mathtools}

\newcommand{\aop}{\hat a}
\newcommand{\nop}{\hat n}
\newcommand{\ntot}{\hat N}
\newcommand{\hop}{\hat H}

\newcommand{\dens}{\hat\rho}
\newcommand{\halfop}{\hat N_A}

\newcommand{\tempflu}{\mathcal{T}}
\newcommand{\cor}{\mathcal{C}}

\let\vec\oldvec
\newcommand{\vec}{\mathbf}
\newcommand{\mat}{\textsf}

\makeatletter
\def\bbl@set@language#1{%
  \edef\languagename{%
    \ifnum\escapechar=\expandafter`\string#1\@empty
    \else\string#1\@empty\fi}%
  \@ifundefined{babel@language@alias@\languagename}{}{%
    \edef\languagename{\@nameuse{babel@language@alias@\languagename}}%
  }%
  \select@language{\languagename}%
  \expandafter\ifx\csname date\languagename\endcsname\relax\else
    \if@filesw
      \protected@write\@auxout{}{\string\select@language{\languagename}}%
      \bbl@for\bbl@tempa\BabelContentsFiles{%
        \addtocontents{\bbl@tempa}{\xstring\select@language{\languagename}}}%
      \bbl@usehooks{write}{}%
    \fi
  \fi}
\newcommand{\DeclareLanguageAlias}[2]{%
  \global\@namedef{babel@language@alias@#1}{#2}%
}
\makeatother
\DeclareLanguageAlias{en}{english}

\makeatletter
\def\MT@warn@unknown{}
\makeatother

\graphicspath{{./figures/}}

\begin{document}

\title{Probing many-body localization phase transition with superconducting circuits}

\author{Tuure Orell$^1$}
\author{Alexios A. Michailidis$^2$}
\author{Maksym Serbyn$^2$}
\author{Matti Silveri$^1$}

\affiliation{$^1$Nano and Molecular Systems Research Unit, University of Oulu, 90014 Oulu, Finland\\
$^2$IST Austria, Am Campus 1, 3400 Klosterneuburg, Austria}

\date{\today}

\begin{abstract}
Chains of superconducting circuit devices provide a natural platform for studies of synthetic bosonic quantum matter. Motivated by the recent experimental progress in realizing disordered and interacting chains of superconducting transmon devices, we study the bosonic many-body localization phase transition using the methods of exact diagonalization as well as matrix product state dynamics. We estimate the location of transition separating the ergodic and the many-body localized phases as a function of the disorder strength and the many-body on-site interaction strength. The main difference between the bosonic model realized by superconducting circuits and similar fermionic model is that the effect of the on-site interaction is stronger due to the possibility of multiple excitations occupying the same site. The phase transition is found to be robust upon including longer-range hopping and interaction terms present in the experiments. Furthermore, we calculate experimentally relevant local observables and show that their temporal fluctuations can be used to distinguish between the dynamics of Anderson insulator, many-body localization, and delocalized phases. While we consider unitary dynamics, neglecting the effects of dissipation, decoherence and measurement back action, the timescales on which the dynamics is unitary are sufficient for observation of characteristic dynamics in the many-body localized phase. Moreover, the experimentally available disorder strength and interactions allow for tuning the many-body localization phase transition, thus making the arrays of superconducting circuit devices  a promising platform for exploring localization physics and phase transition.
\end{abstract}
\maketitle

\section{Introduction}
\begin{figure}[t]
    \centering
    \includegraphics[width=1.0\linewidth]{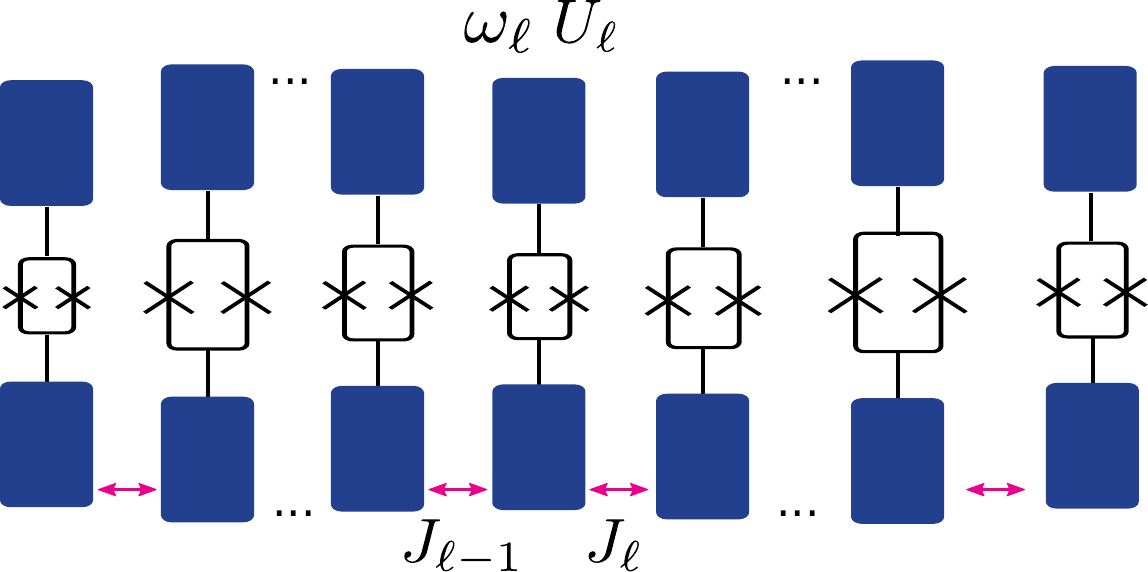}
    \caption{Schematic of a transmon chain realizing the Bose--Hubbard model with attractive interactions. A transmon is made of Josephson junctions (black crosses) and capacitor plates (blue rectangles). Transmons are anharmonic oscillators with natural frequency $\omega$ and anharmonicity~$U$ and they interact with each other through capacitive interaction~$J$. Non-identical sites visualize fabrication scatter and disorder. }
    \label{fig:1}
\end{figure}

Superconducting circuits, specifically arrays of superconducting transmon devices (Fig.~\ref{fig:1}), are a promising platform for quantum simulation. Transmon is a weakly anharmonic oscillator made of Josephson junctions and capacitors. However, scatter in circuit parameters, such as on-site energy and interaction strengths, is inevitable in the process of fabrication and hinders their applications~\cite{Underwood12}. Parameters of different circuit elements can be made also \textit{in situ} tunable with microwave and magnetic flux controls, and disorder can be removed. This tunability requires extra junctions~\cite{kochChargeinsensitiveQubitDesign2007}, engineered couplers~\cite{Chen14} and control lines, which all adds to device, characterization and measurement complexity. Moreover, these ingredients add decoherence and dissipation and thus are not the most advantageous methods for large arrays. 

Presence of intrinsic disorder suggests that transmon arrays can be a natural platform to study physics emergent from interplay between disorder and interactions. Sufficiently strong disorder may lead to a \emph{many-body localized phase} --- a stable phase of matter characterized by the breakdown of thermalization~\cite{nandkishoreManyBodyLocalizationThermalization2015, Alet18, Abanin18}. Many-body localized systems are characterized by a logarithmically slow entanglement spreading~\cite{serbynSlowGrowthEntanglement} and relaxation to a non-thermal state that retains memory of initial conditions. This is in contrast to ballistic spreading of entanglement in thermalizing systems that achieve thermal equilibrium and lose memory of the initial state~\cite{ChaosReview}. The characteristic slow dynamics of the many-body localized phase makes it a prospective target for quantum simulators that are characterized by slow loss of coherence~\cite{Choi17, Zhang17, Sacha18}.

Many-body localization is an active research field at the intersection of non-equilibrium quantum dynamics, quantum thermalization, condensed matter physics, quantum information, computational physics, and other fields~\cite{nandkishoreManyBodyLocalizationThermalization2015, Alet18, Abanin18}. The major focus of earlier studies was on localization in one dimensional systems with the two-dimensional local Hilbert space. Specific examples of such systems include interacting spinless fermions~\cite{oganesyan07, barlev14, barlev2015}, spin-1/2 chains~\cite{luitzManybodyLocalizationEdge2015, singhSignaturesManybodyLocalization2016, pal10, nanduriEntanglementSpreading, znidaricManybodyLocalization}, and hard-core bosons~\cite{santos10} on one-dimensional lattices. In contrast, localization in bosonic systems received relatively little attention, with an exception of a few works~\cite{Aleiner10, Sierant2017, sierantManybodyLocalizationBosons2018a, sierantManybodyLocalizationPresence2019}. The bosonic systems are more challenging for numerical studies~\cite{sierantManybodyLocalizationBosons2018a}, since the size of the local Hilbert space in  a bosonic model with particle conservation is limited only by the total number of excitations in the system.  This implies that the scaling of the total Hilbert space with system size heavily depends on the filling factor. Furthermore, as we show below, the disorder-dependent asymmetry in the density of states presents additional complications.

Recent research considered interacting microwave photons in disordered~\cite{roushanSpectroscopicSignaturesLocalization2017, Neill17} and clean~\cite{hacohen-gourgyCoolingAutonomousFeedback2015, maDissipativelyStabilizedMott2019} superconducting circuit chains as well as bosonic Rb-atoms in optical lattices~\cite{Choi16,Rispoli18, lukinProbingEntanglementManybody2019}, and demonstrated that bosonic systems provide good experimental platforms to study many-body physics. Motivated by this experimental progress with bosonic synthetic matter, we study here a chain of superconducting transmon devices (Fig.~\ref{fig:1}) realizing a disordered Bose--Hubbard model with attractive interactions~\cite{hacohen-gourgyCoolingAutonomousFeedback2015, Kirchmair15} (Sec.~\ref{sec:transarray}). 

After introducing the model, we proceed with numerical calculation of the phase diagram, identifying the location of the phase transition that separates ergodic and many-body localized phases as a function of disorder strength and interactions (Sec.~\ref{sec:phase}). In addition, we  consider experimentally relevant longer-range hopping and interaction terms in the Hamiltonian and study their effect on the many-body localization phase transition. Crucially, we demonstrate that the phase transition stays within the experimentally accessible range of parameters in the transmon chains. 

In the second half of the paper, we consider time dynamics of local observables after a quench from an initial product state (Sec.~\ref{sec:dyn}). We show that the fluctuations of local observables can be used to experimentally distinguish between the dynamics of Anderson insulator, many-body localization and delocalized phases on the experimentally accessible time scales. In other words, temporal fluctuations of local observables can serve as an alternative to the challenging measurement of the entanglement dynamics in a quantum quench~\cite{lukinProbingEntanglementManybody2019, brydgesProbingRenyiEntropy}. 
Finally, we conclude with discussing future directions, and possible effects arising from dephasing, dissipation and measurement back action (Sec.~\ref{sec:conc}).

In addition to our two main results, the phase diagram of the bosonic Hubbard model and dynamics of local observables, our work extends the numerical methods for studies of many-body localization to bosonic systems. We utilize methods of exact diagonalization for solving eigenstates and resolving the phase diagram. Dynamics are performed using Krylov subspace methods~\cite{luitzErgodicSideManybody} as well as time evolving block decimation~\cite{vidal2004efficient}, which is a Trotter-based evolution scheme for matrix product states~\cite{pirvu2010matrix}. As bosonic systems are rather an untrodden path in the context of numerics of interacting and disordered systems, we also present details on the used numerical methods and their scalability (App.~\ref{app:ex}-\ref{app:timeMPS}). In particular, we discuss the combination of the~$LDL$~decomposition and the stochastic Chebyshev series expansion~\cite{napoliEfficientEstimationEigenvalue2016} for efficient estimation of the density of states without solving the full energy eigenvalue spectrum. 

\section{Transmon array}\label{sec:transarray}
Transmon is a weakly anharmonic electric oscillator with natural frequency $\omega=\sqrt{8E_{\rm C}E_{\rm \Sigma J}}/\hbar$ made of capacitor plates setting the capacitive charging energy $E_{\rm C}$ and Josephson junctions acting as a weakly non-linear inductor~\cite{kochChargeinsensitiveQubitDesign2007, Paik11} setting the total Josephson energy $E_{\rm \Sigma J}$. In the transmon regime, the Josephson energy dominates over the charging energy, characterized by the ratio $E_{\rm \Sigma J}/E_{\rm C}$ in the range of \numrange{25}{100}. In this work, we are interested in probing many-body dynamics in a chain of transmons. The transmon chain is schematically depicted in Fig.~\ref{fig:1} assuming  transmons with a 3D architecture~\cite{Paik11, Kirchmair15} placed in a 3D cavity or a waveguide (not shown). 3D architecture offers benefits in terms of low dissipation and decoherence rates as well as versatile geometric options in coupling of the transmons. In contrast, transmons with a 2D architecture coupled to 2D cavities made on coplanar waveguide resonators~\cite{roushanSpectroscopicSignaturesLocalization2017, maDissipativelyStabilizedMott2019} have very accurate, fast and developed measurement and controlling schemes. However, both architectures are essentially similar in the point of view of unitary many-body dynamics.

Given the recent experimental demonstrations with chains and clusters of \numrange{10}{20} superconducting qubits~\cite{roushanSpectroscopicSignaturesLocalization2017, Xu17, maDissipativelyStabilizedMott2019, Song19}, the arrays are expected to reach \numrange{10}{100} sites in the near future. Transmons interact with each other via capacitive dipole-dipole interaction~$J$, whose strength can be tailored from \SI{10}{\mega\hertz} to \SI{100}{\mega\hertz} by changing the orientation and size of  transmon capacitors~\cite{hacohen-gourgyCoolingAutonomousFeedback2015,Kirchmair15}. The interaction strength should be compared with the dissipation and decoherence rates of the state-of-the-art devices~\cite{Chang13, Martinis14, Paik11, rigetti2012, Wendin2017, gambettaBuildingLogicalQubits2017, peterer2015}. The decoherence rates~$\Gamma_2$ range from~\SI{10}{\kilo\hertz} in 3D architecture transmons to~\SI{25}{\kilo\hertz} in 2D architecture transmons. Similarly, the dissipation rates~$\Gamma_1$ range from~\SI{2}{\kilo\hertz} to~\SI{5}{\kilo\hertz} in 3D and 2D architectures, respectively. Almost three orders of magnitude difference between the interaction strength and dissipation/decoherence rates yields an ample time frame of unitary many-body dynamics before disruptive dissipation and decoherence effects become important. 

In quantum computing applications one uses the transmon anharmonicity~$U=E_{\rm C}/\hbar$ in the order of a few \SI{100}{\mega\hertz}~\cite{Koch10,Paik11} to turn transmons into effective two-level systems.~\cite{Devoret13} However, in the quantum simulations of the many-body physics, the higher levels and their bosonic excitation statistics can be accessed. In this case, the anharmonicity acts as an on-site interaction between bosonic excitations. Superconducting transmon devices can be easily driven with single-site accuracy offering a possibility to study driven-dissipative dynamics~\cite{Houck12}. Furthermore, superconducting qubit read-out can be made almost perfectly quantum non-demolition~\cite{Sun14, Ofek16} offering a possibility to explore many-body dynamics and entanglement phase transitions~\cite{Skinner18, Li18, Chan18, Li19, Choi19, GullansHuse19} under continuous repeated or variable strength~\cite{Hatridge13} measurements.

In this Section, we introduce the non-disordered Bose--Hubbard model realized by a chain of identical trasmons. In addition we discuss additional terms present in the Hamiltonian --- the longer range and higher-order multi-particle interactions as well as transmon-specific disorder potential. As our main focus lies in many-body localization induced by the presence of strong disorder, we continue by discussing different ways to realize in-situ tunable disorder and calculate the resulting many-body eigenspectrum. 

\subsection{Clean Bose--Hubbard model}
One-dimensional array of $L$ identical transmons (Fig.~\ref{fig:1}) is described by the attractive Bose--Hubbard model~\cite{maDissipativelyStabilizedMott2019, roushanSpectroscopicSignaturesLocalization2017, hacohen-gourgyCoolingAutonomousFeedback2015}
\begin{align}
    \hop_{\rm BH}/\hbar =\sum_{\ell=1}^L \omega \nop_\ell -
    &\sum_{\ell=1}^L\frac{U}{2}\nop_\ell\left(\nop_\ell - 1\right) \notag \\ 
    +&\sum_{\ell=1}^{L-1}J\left(\aop_{\ell}^\dagger\aop^{}_{\ell+1} 
    + \aop^{}_{\ell}\aop_{\ell+1}^\dagger\right),\label{eq:2_1}
\end{align}
where~$\aop^{}_\ell$ and~$\aop_\ell^\dagger$ are the bosonic annihilation and creation operators of the site~$\ell$, $[\aop^{}_\ell, \aop_k^\dag]=\delta_{\ell,k}$, and~$\nop_\ell = \aop_\ell^\dagger\aop^{}_\ell$ is the corresponding number operator. In the many-body language, the natural frequency~$\omega$ corresponds to the on-site energy, capacitive dipole-dipole interaction $J$ is interpreted as the nearest-neighbor hopping between adjacent sites, and the anharmonicity $U=E_{\rm C}/\hbar$ serves as the on-site interaction. The Hamiltonian~\eqref{eq:2_1} conserves the total number of excitations since it commutes with the total occupation operator~$\ntot = \sum_{\ell=1}^L\nop_\ell$.  It is therefore sufficient to study only a single subspace of the Hamiltonian with a fixed total number of excitations~$N$. Here we focus mainly on the chain at half-filling, i.e.~$N=L/2$, with filling factor $f=N/L=1/2$. The experimentally relevant parameters are summarized in Table~\ref{tab:2_1}.

We note that the on-site interaction is attractive. It stems from the cosine potential of the Josephson junctions, $E_{\rm J}\cos[\sqrt[4]{2E_{\rm C}/E_{\Sigma\rm J}}(\hat a+\hat a^\dag)]$, that softens as a function of the excitation number. The  anharmonic term $U=E_{\rm C}/\hbar$ is the lowest order correction to the harmonic potential and  as the occupation number increases, one has to take into account also higher order corrections, the first of such being
\begin{equation}\label{eq:2_3}
    \hop_{\rm HA}/\hbar =\sum_{\ell=1}^L \frac{U_2}{6}
    \nop_\ell\left(\nop_\ell - 1\right)\left(\nop_\ell - 2\right).
\end{equation}
Here, the higher-order anharmonicity, which is repulsive with $U_2/2\pi$ in the range \SIrange{10}{30}{\mega\hertz}, effectively reduces total anharmonicity of the transmon. The cosine potential also implies that there exists a finite number of bound states~\cite{Leghtas1, Leghtas2, pietikainenQuantumtoclassicalTransitionDrivendissipative2019}, denoted as the transmon states, whose total number per site is approximately $\sqrt{E_{\rm J}/E_{\rm C}}$. Because the excitations in this system are bosons, there is a possibility that they all occupy the same site.  These facts introduce a theoretical upper limit for the validity of our model. Depending on the parameters, only~$\sqrt{E_{\rm J}/E_{\rm C}}\sim 10$ lowest states of each transmon are bound. Higher occupations on single sites break the Bose--Hubbard approximation limiting the validity of our model to~$\sim 20$ transmons at half-filling. 

In addition to the nearest-neighbor interaction, in realistic systems there exists also longer range tunneling process between sites further apart. Tunneling between next-nearest neighbors is described by term
\begin{equation}\label{eq:2_2}
    \hop_{\rm LR}/\hbar = \sum_{\ell=1}^{L-2}J_2
    \left(\aop_{\ell}^\dagger\aop^{}_{\ell+2} + \aop^{}_{\ell}\aop_{\ell+2}^\dagger\right).
\end{equation}
The next-nearest neighbor hopping is typically weak compared to the nearest neighbor hopping,~$J_2\sim J/10$~\cite{hacohen-gourgyCoolingAutonomousFeedback2015}. Notice that the additional perturbations of Eqs.~\eqref{eq:2_3} and~\eqref{eq:2_2} also conserve the total number of excitations. 

\begin{table}[t]
    \centering
    \begin{tabular}{c|c|c}
       Parameter & Symbol & Value \\
       \hline
        on-site energy &  $\omega/2\pi$ & \SIrange{5}{10}{\giga\hertz} \\
      on-site interaction & $U/2\pi$ & \SIrange{200}{300}{\mega\hertz}\\
      hopping  & $J/2\pi$ & \SIrange{10}{100}{\mega\hertz}\\
      disorder amplitude & $W/2\pi$ & \SI{10}{\mega\hertz}--\SI{2}{\giga\hertz}\\
     transmon asymmetry  & $d$ & 0.1
    \end{tabular}
     \caption{Experimentally relevant values for the parameters in the disordered Bose--Hubbard Hamiltonian~\eqref{eq:2_6}. The dimensionless ratios are in the range $U/J = 2 - 30$ and $W/J = 0.1 - 200$~\cite{maDissipativelyStabilizedMott2019, roushanSpectroscopicSignaturesLocalization2017, hacohen-gourgyCoolingAutonomousFeedback2015}, which we demonstrate to be sufficient for tuning the many-body localization phase transition.}
     \label{tab:2_1}
\end{table}

\begin{figure}
    \centering
    \includegraphics[width=1.0\linewidth]{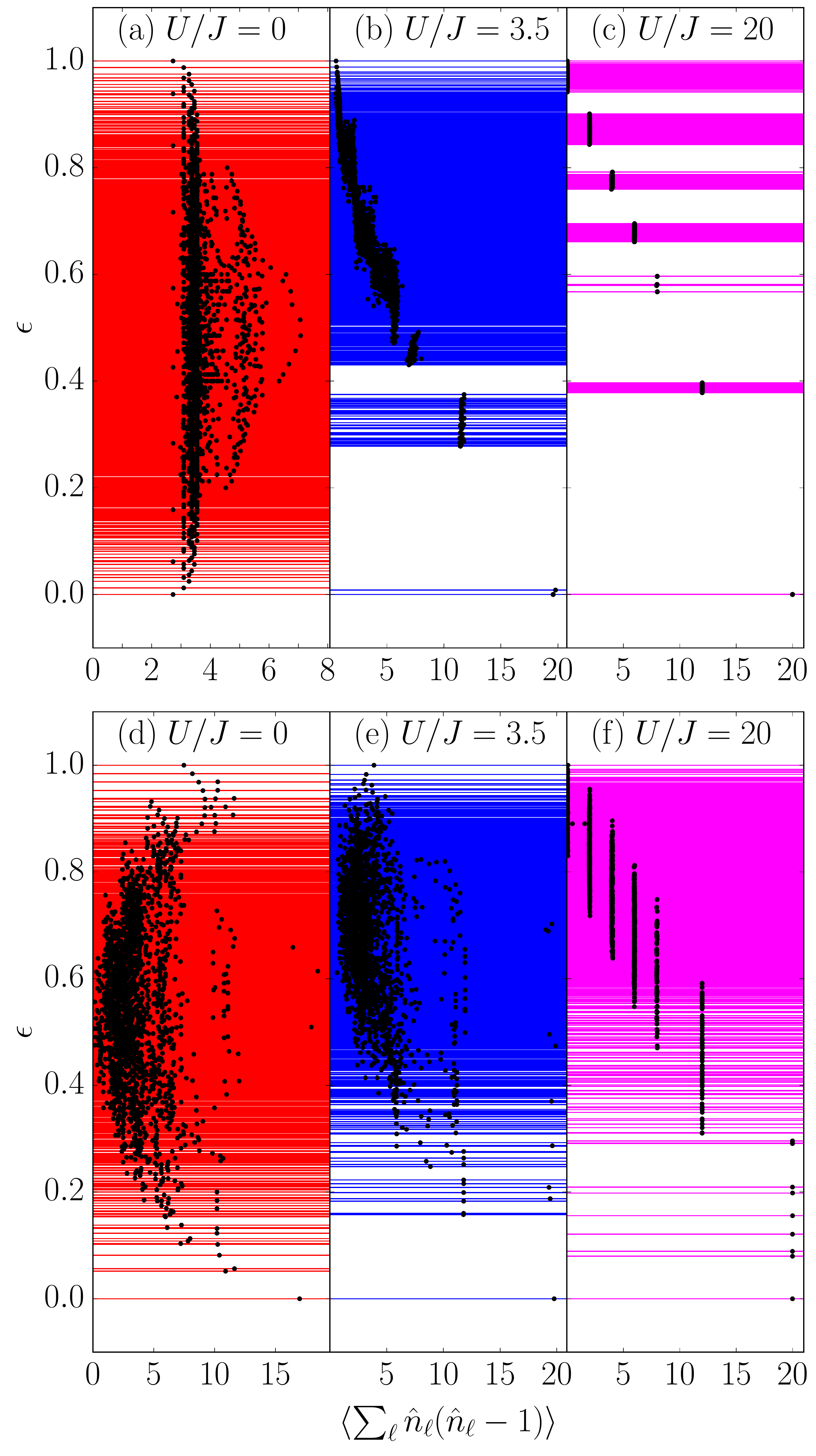}
    \caption{Normalized eigenenergies of the clean (a)-(c) and the disordered (d)-(f) Bose--Hubbard Hamiltonian at half-filling and $L=10$ for anharmonicities $U/J=0$ (a) and (d); $U/J=3.5$ (b) and (e); $U/J=20$ (c) and (f). The eigenenergies are scaled as $\epsilon = (E-E_{\rm min})/(E_{\rm max}-E_{\rm min})$. The disorder realization, drawn from a uniform distribution with the disorder strength $W/J=10$, is the same for each $U$. Black dots show the expectation value of the total anharmonicity operator $\sum_\ell\nop_\ell(\nop_\ell-1)$ in the corresponding eigenstate.}
    \label{fig:2_1}
\end{figure}

At zero-temperature the clean Bose--Hubbard Hamiltonian~\eqref{eq:2_1} can undergo a phase transition between the Mott-insulating and superfluid phases~\cite{fisherBosonLocalization1989}. This is a ground state phase transition studied in repulsive Bose--Hubbard model. In this work, we are interested in the highly excited or infinite-temperature eigenstates of the attractive Bose--Hubbard model. Therefore it is important to understand the structure of the many-body eigenspectrum and especially how it is affected by the on-site interaction~$U$. We show the full energy spectrum of Eq.~\eqref{eq:2_1} with~$L=10$ at half-filling in Fig.~\ref{fig:2_1} for three values of the on-site interaction $U/J=0$~(a), $3.5$~(b) and $20$~(c). When discussing the energies of  excited many-body eigenstates (shown in Fig.~\ref{fig:2_1} as horizontal lines), it is often convenient to consider the normalized energy 
\begin{equation}
    \epsilon = \frac{E - E_{\rm min}}{E_{\rm max} - E_{\rm min}} \in [0, 1],
\end{equation}
where~$E_{\rm min}$ and~$E_{\rm max}$ are the smallest and largest eigenvalues of the Hamiltonian in the studied sector and $E \in [E_{\rm min}, E_{\rm max}]$ is an arbitrary energy eigenvalue.

In the absence of on-site interaction $U$,  the Bose--Hubbard Hamiltonian describes a chain of coupled harmonic oscillators. This system has a symmetric spectrum [Fig.~\ref{fig:2_1}(a)], but as the anharmonicity is increased in Figs.~\ref{fig:2_1}(b)-(c), the symmetry is removed and the eigenstates begin to form mini-bands. Because of the negative anharmonicity, two excitations occupying the same transmon have smaller energy than two excitations on different sites. The lowest energies are obtained when all excitations occupy the same transmon. This is seen in the expectation values of the total anharmonicity operator~$\sum_\ell\nop_\ell(\nop_\ell - 1)$ (black dots in Fig.~\ref{fig:2_1}): the larger the expectation value, the more bosons are occupying a single site. This behavior is visible in Fig.~\ref{fig:2_1}(b)-(c) showing that in systems with non-zero~$U$ the expectation value decreases as the energy is increased. In the limit of large~$U/J$, the mini-bands are fully formed, the total anharmonicity is conserved within each band separately and the nearest-neighbor hopping interaction weakly lifts the degeneracy within the bands.

To calculate the many-body spectrum, we utilize exact diagonalization of Hamiltonian~\eqref{eq:2_1}. Exact numerical calculations of many-body quantum systems are very demanding due to the exponential scaling of the Hilbert space dimension. In our case we are interested in~$L$ coupled~$N+1$-level systems with a total Hilbert space dimension~$D_L = (N+1)^L$. Due to the conservation of total number of quanta, Hamiltonian is a block-diagonal matrix where each block is characterized by a total number of excitations~$\braket\ntot = N$. In a system with~$L$ sites these individual blocks have dimensions
\begin{equation}\label{eq:2_states}
    D_{N,L} = \frac{(N + L - 1)!}{(L-1)!N!} 
    \approx \frac{\left[\left(1+f\right)\left(1+f^{-1}\right)^f\right]^L}{\sqrt{L}},
\end{equation}
where the last form is an approximation in the long-chain limit with filling factor~$f = N/L$. For a spin chain the size of a sector with zero total magnetization (the largest sector) scales approximately as~$2^L/\sqrt{L}$~\cite{pietracaprinaShiftinvertDiagonalizationLarge2018a}. In the long-chain limit, the Bose--Hubbard chain with half-filling scales roughly as~$2.6^L/\sqrt{L}$, and with unit filling already as~$4^L/\sqrt{L}$. This naturally implies that we cannot reach as large system sizes as used in spin chain studies, where the currently reached upper limit is~$L=24$~\cite{pietracaprinaShiftinvertDiagonalizationLarge2018a} for exact numerical studies of the Hamiltonian eigenstates. The largest spin chain studies rely on massive parallelization on distributed memory machines. Our simulations use smaller computational resources, but we are still able to study systems with~$14$ transmons, which have Hilbert space roughly comparable to that of~spin-1/2 chain with~$18$ sites. The chain with~$16$ transmons is already comparable to~$22$ spins, so it seems to be the current practical upper limit for this kind of study.

Technically, in a sector with~$N$ quanta each transmon needs~$N+1$ lowest energy levels. One can then construct all the necessary operators in this desired sector. This can be done by first building all the possible Fock state configurations in this sector, i.e. states~$\ket{n_1,n_2,n_3,\dots,n_L}$ with $\sum_{\ell=1}^Ln_\ell = N$. One can then obtain the necessary matrix elements by considering how the individual operators such as~$\aop_1^\dagger\aop_2$ should operate on these basis states. The operators are most efficiently implemented as sparse matrices, which is done here with Eigen-library~\cite{eigenweb} for C++. The resulting Hamiltonian is a sparse matrix. However, we note that the next-nearest neighbor interaction of Hamiltonian~\eqref{eq:2_2} reduces the sparseness and makes exact diagonalization numerically more demanding problem than the regular Bose--Hubbard model. For more details of the exact diagonalization, see App.~\ref{app:ex}.

\subsection{Flux-tunable disorder potential}
\begin{figure}
    \centering
    \includegraphics[width=1.0\linewidth]{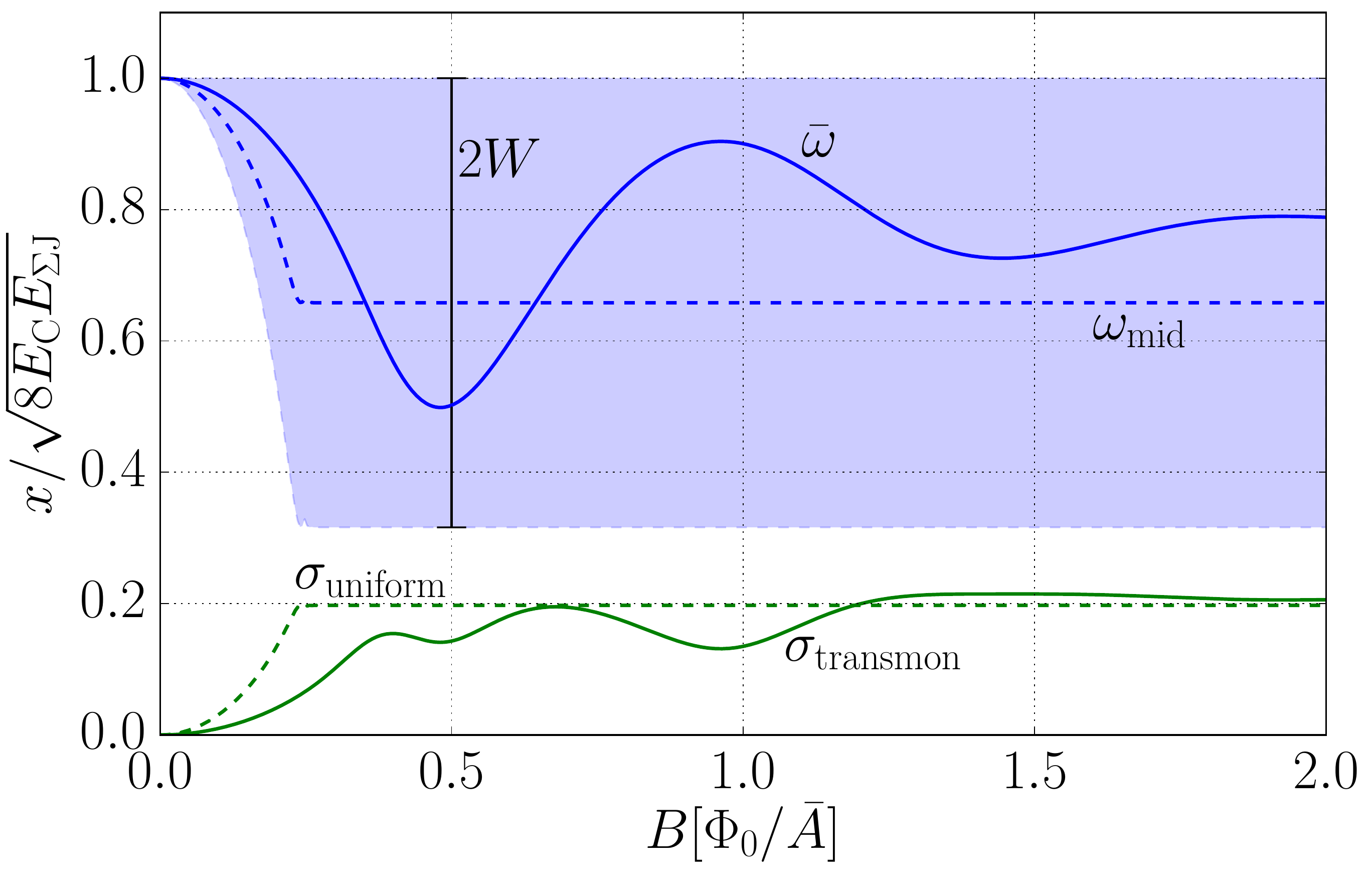}
    \caption{Behavior of the mean energy~$\bar\omega$ and corresponding standard deviation~$\sigma_{\rm transmon}$ of the transmon potential as a function of magnetic field.
    The fact that the mean energy (blue curve) is close to the ``middle'' energy $\omega_\text{mid} = (\omega_\text{max}+\omega_\text{min})/2$ (blue dashed curve) at high magnetic fields suggests that the uniform distribution provides a good approximation to the true distribution of $\omega_l$. The standard deviation (green curve) of the transmon disorder potential of Eq.~\eqref{eq:2_5} as a function of magnetic field also becomes close to the one of the uniform distribution~$\sigma_{\rm uniform} = W/\sqrt{3}$ (dashed green curve) for strong magnetic field. In contrast, weak magnetic fields give the narrow distribution of transmon energies sharply peaked around the mean energy. Junction asymmetry is~$d = 0.1$,  loop-areas are assumed to be Gaussian distributed with standard deviation~$1/5$ of the mean $\bar A$ and all the values are expressed in the units of unperturbed transmon energy~$\sqrt{8E_{\rm C}E_{\Sigma\rm{J}}}$.}
    \label{fig:2_0}
\end{figure}

Since our goal is to study the many-body localization phase transition tuned by disorder, we explore the means to control disorder amplitude in situ. With transmons this is possible via overall magnetic flux tuning~\cite{Fitzpatrick17}. A single transmon can be made to consist of two parallel Josephson junctions with energies~$E_{\rm J1}$ and~$E_{\rm J2}$~\cite{kochChargeinsensitiveQubitDesign2007} connected by a loop with a surface area~$A$, see Fig.~\ref{fig:1}. In this case the on-site energy of a transmon depends on the magnetic flux~$\Phi = BA$ induced by a uniform magnetic field~$B$ threading the loop. Thus, the on-site energy of a transmon is~\cite{kochChargeinsensitiveQubitDesign2007}
\begin{equation}
    \omega(\Phi) = \frac{\sqrt{8E_{\rm C}E_{\Sigma\rm J}}}{\hbar}
    \sqrt[4]{\cos^2\left(\frac{\pi\Phi}{\Phi_0}\right) +
    d^2\sin^2\left(\frac{\pi\Phi}{\Phi_0}\right)},
\end{equation}
where $\Phi_0 = h/2e$ is the superconducting flux quantum, $E_{\Sigma\rm J} = E_{\rm J1} + E_{\rm J2}$ is the sum of the junction energies and $d = \frac{E_{\rm J1} - E_{\rm J2}}{E_{\rm J1} + E_{\rm J2}}\sim\pm 0.1$ is the junction asymmetry.

The process of multiple transmon fabrication typically results in a Gaussian distributed variation of the loop areas $A_\ell$ of the resulting devices. If an array constructed from these transmons is placed in the external magnetic field, a Gaussian distributed magnetic flux~$\Phi_\ell = BA_\ell$ is induced in each transmon, which leads to a non-uniformly distributed on-site energies $\omega_{\ell}$,
\begin{equation}\label{eq:2_5}
    \omega_\ell(B) =\frac{\sqrt{8E_{\rm C}E_{\Sigma\rm J}}}{\hbar}
    \sqrt[4]{\cos^2\left(\frac{\pi BA_\ell}{\Phi_0}\right) +
    d^2\sin^2\left(\frac{\pi BA_\ell}{\Phi_0}\right)}. 
\end{equation}
The energy is bounded above by $\omega_{\rm max}=\sqrt{8E_{\rm C}E_{\rm \Sigma J}}/\hbar$ and the junction asymmetry sets the lower bound $\omega_{\rm min}=\sqrt{d}\omega_{\rm max}$. Since the energy depends non-linearly on the magnetic flux, both the mean energy and the variance of the non-uniform energy distribution of Eq.~\eqref{eq:2_5} can be in-situ controlled with the uniform magnetic field $B$, as visualized in Fig.~\ref{fig:2_0}.

The resulting distribution of $\omega_{\ell}$ is non-uniform, so that in general the mean energy $\bar\omega$ differs from the average of largest and smallest energy. Nevertheless, the mean energy is the same for each transmon and has no effect on the many-body dynamics since it can be removed by switching to a rotating coordinate system with $\hat U=\exp(-i \bar{\omega} t \sum_\ell  \nop_\ell)$. The random energies are distributed around their mean value with width $2W = \omega_{\rm max} - \omega_{\rm min}$, see Fig.~\ref{fig:2_0}. Thus, we can use the parameter~$W$ as an approximate disorder strength and assume that the distribution is uniform,~$\omega_\ell \sim [-W, W]$ with the resulting standard deviation~$W/\sqrt{3}$. 

In Fig.~\ref{fig:2_0} we show that the disorder strength~$W$ exhibits non-linear growth at weak magnetic field, but it quickly saturates due to the saturation of the minimum energy to value~$\omega_{\rm min} =\sqrt{d}\sqrt{8E_{\rm C}E_{\Sigma \rm J}}$.  Notice that because loop area is neither correlated with the total Josephson energy nor the charging energy, in the absence of magnetic field and assuming no fabrication disorder in junctions, all the transmons are nominally identical. Weak fabrication disorder~\cite{Kirchmair15} breaks this by inducing a lower bound for the attainable experimental disorder $W_{\rm min}/2\pi\approx$~\SI{10}{\mega\hertz}. In practice, however, the exact form for the disorder potential influences only non-universal details, such as the exact location of the phase transition. Therefore, in what follows, we use the uniform distribution instead of the transmon potential due to its simpler form and to facilitate the comparison with other studies of many-body localization~\cite{luitzManybodyLocalizationEdge2015, singhSignaturesManybodyLocalization2016, sierantManybodyLocalizationBosons2018a, mondainiManybodyLocalizationThermalization2015, pal10}.

\subsection{Disordered Bose--Hubbard model}
As discussed above, the on-site energy can be made strongly disordered by flux tuning. In addition, anharmonicity and tunneling terms can also contain disorder through fabrication and dependence on the flux-tunable Josephson energy. We can write the Hamiltonian of the disordered Bose--Hubbard model as
\begin{align}
    \hop/\hbar = \sum_{\ell=1}^L\omega_\ell\nop_\ell 
    -&\sum_{\ell=1}^L \frac{U_\ell}{2}\nop_\ell\left(\nop_\ell - 1\right) \notag \\
    +&\sum_{\ell=1}^{L-1}J_\ell\left(\aop_{\ell}^\dagger\aop^{}_{\ell+1} 
    + \aop^{}_{\ell}\aop_{\ell+1}^\dagger\right), \label{eq:2_6}
\end{align}
and we can also include disordered higher-order anharmonicity and next-nearest neighbor hopping in Eqs.~\eqref{eq:2_3} and~\eqref{eq:2_2}. Experimentally achievable parameters for this Hamiltonian are listed in Table~\ref{tab:2_1}. The disorder in hopping and on-site interaction can be drawn e.g. from a Gaussian distribution. Because the disorder in the on-site energy reaches much larger values than is possible for the hopping and on-site interaction, we mainly focus on the situation where disorder is included only in the on-site energies~$\omega_\ell$ of Eq.~\eqref{eq:2_6}.

\begin{figure}
    \centering
    \includegraphics[width=1.0\linewidth]{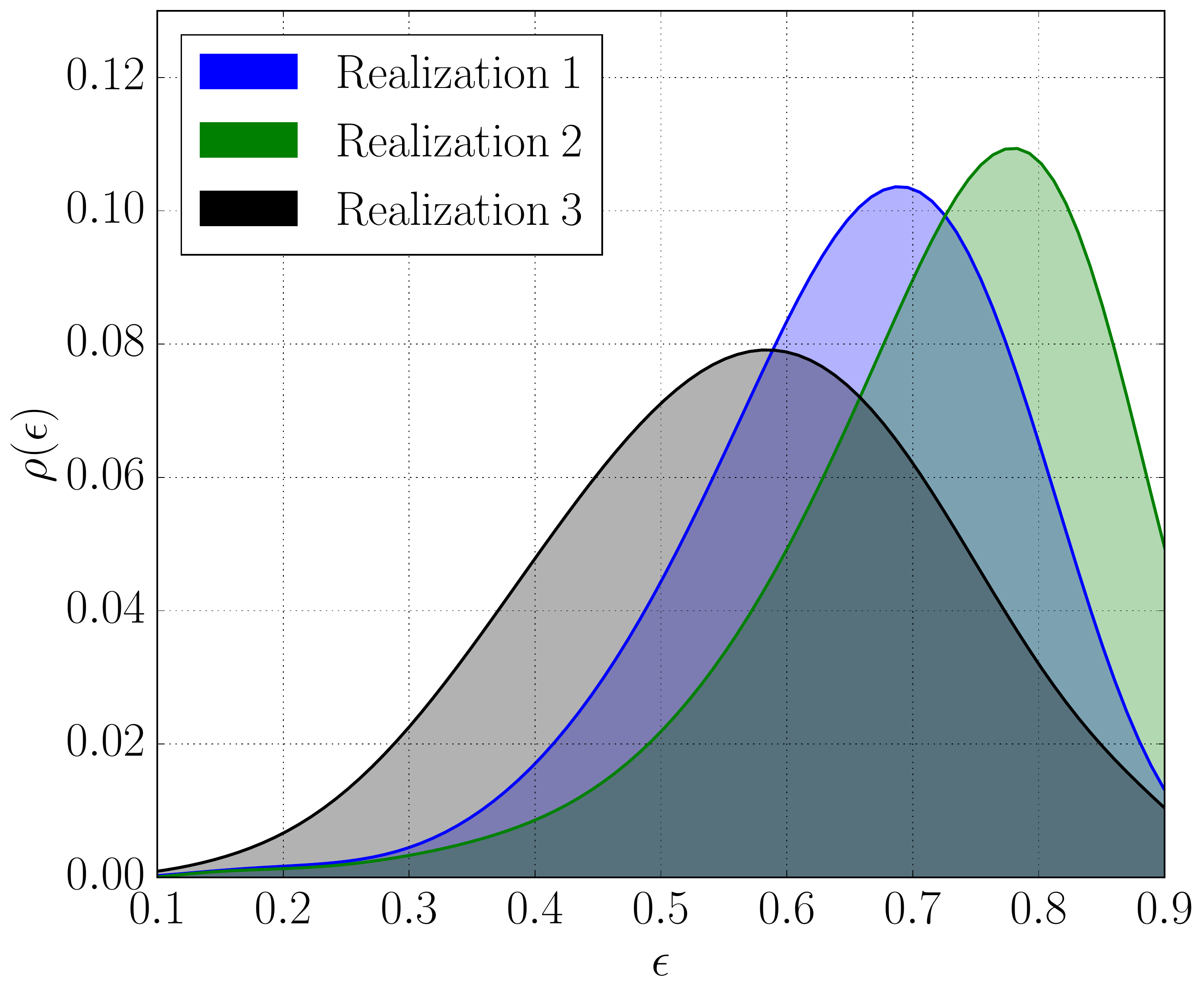}
    \caption{The density of states $\rho(\epsilon)$ in the disordered Bose--Hubbard model of Eq.~\eqref{eq:2_6} as a function of normalized energy~$\epsilon$ for three different disorder realizations with parameters $L=10$, $U/J = 3.5$ and $W/J = 10$. The location of the maximum varies due to interplay of anharmonicity and disorder. The realization 1 (blue) is the same as in Fig.~\ref{fig:2_1}(e).}
    \label{fig:2_3}
\end{figure}

The many-body eigenenergies of the disordered Bose--Hubbard model are shown in Figs.~\ref{fig:2_1}(d)-(f). We observe that the disorder changes the structure of the spectrum from that of the clean system shown in Figs.~\ref{fig:2_1}(a)-(c). Because sites are no longer identical, there exists a preferred site with the lowest on-site energy and the configuration where all excitations occupy this site gives a good approximation to the ground state of the system. For sufficiently weak anharmonicity the disorder dominates and the expectation value of the total anharmonicity operator can be large even in high-energy eigenstates. For strong anharmonicity, mini bands of the clean system start to overlap, and the clean system structure, where the expectation value of the total anharmonicity operator decreases with increasing energy, still remains. For large disorder the eigenstates are approximately the Fock states. 

\subsection{Density of states}
The interplay of disorder and anharmonicity also has impact on the density of states, which varies between different disorder realizations, as shown in Fig.~\ref{fig:2_3}. Especially we note that the normalized energy at which the density of states has a maximum depends on the realization, as well as on the disorder strength~$W$ and the on-site interaction~$U$. The same behavior is observed also with the density of states plotted as a function of energy $E$ rather than rescaled energy $\epsilon$.

Since many-body localization is a property of highly-excited eigenstates, the disorder-dependent density of state introduces an additional complexity. In the most studied model of the many-body localization, the Heisenberg spin chain~\cite{luitzManybodyLocalizationEdge2015, singhSignaturesManybodyLocalization2016}, the spectrum is symmetric and the density of states has much weaker dependence on disorder realizations with a maximum located in the middle of the spectrum~($\epsilon = 0.5$). Thus, in that model one can choose some fixed normalized energy and compare the corresponding many-body eigenstates and their properties (bipartite entanglement entropy, bipartite number uncertainty etc.) between different disorder realizations and eventually average the properties over a narrow energy window around the fixed normalized energy~\cite{luitzManybodyLocalizationEdge2015}. However, for the disordered Bose--Hubbard model with sufficiently large many-particle interaction strength $U$, this is not possible because the properties of eigenstates obtained this way would vary too much. For example, in one realization the many-body eigenstate closest to the fixed normalized energy $\epsilon = 0.6$ might be close to the maximum density of states~[see realization $3$ (black) in Fig.~\ref{fig:2_3}] and thus exhibit infinite-temperature behavior, but in another~[realization 2 (green) in Fig.~\ref{fig:2_3}] it might correspond to one of the low-lying eigenstates. Thus, in order to obtain comparable eigenstates we instead choose them at the maximum density of states of each individual realization. This is physically justified choice since in the Heisenberg spin chain the many-body localization transition is known to require largest disorder strength in eigenstates located at the maximum density of states, due to the influence of the neighboring states which gives rise to the many-body mobility edge~\cite{luitzManybodyLocalizationEdge2015}. Similar behavior is expected also in the disordered Bose--Hubbard model of Eq.~\eqref{eq:2_6}. Downside is that we need at least an estimate for the density of states, which introduces additional numerical complexity.

An estimate of the density of states, without solving the full spectrum of many-body energy levels, can be obtained by using Sylvester's law of inertia~\cite{napoliEfficientEstimationEigenvalue2016}, which requires computation of~$LDL$ decompositions of the Hamiltonian of Eq.~\eqref{eq:2_6}. This results in an exact number of energy eigenvalues within a specified energy interval. If the interval size is sufficiently large, this method produces the density of states more efficiently than the full exact eigendecomposition. Since we are not interested in the full shape of the density of states, but just the location of the maximum, the $LDL$ decomposition based method is very efficient. For large systems, however, the~$LDL$ decomposition [essentially scaling similarly as eigendecomposition, $\mathcal{O}(n^3)$,  where $n$ is the matrix dimension, but with a smaller prefactor] becomes numerically too heavy. Luckily, accurate and efficient approximation of the number of eigenvalues within a specified energy interval, and hence the density of states, can still be obtained with the stochastic Chebyshev expansion method~\cite{napoliEfficientEstimationEigenvalue2016}. This method is efficient for large systems for two main reasons. First, as a stochastic method, speed can be traded to accuracy similar to Monte Carlo methods. Second, important for large systems, this method is based only on sparse~matrix-vector multiplications instead of matrix decomposition. The details of Sylvester's law of inertia,~$LDL$ decomposition, and the stochastic Chebyshev expansion method are presented in App.~\ref{app:LDL}.

\section{Phase transition} \label{sec:phase}

Generic, isolated, and interacting many-body quantum systems reach thermal equilibrium in course of their unitary dynamics~\cite{ChaosReview}. The eigenstate thermalization hypothesis (ETH) provides a microscopic mechanism for thermalization, by imposing the condition that individual eigenstates of the many-body system have thermal expectation values of all local observables~\cite{DeutschETH,SrednickiETH}. Eigenstate thermalization hypothesis has numerous implications for the structure of eigenstates, in particular it suggests a volume-law entanglement in highly excited eigenstates. 

However, the sufficiently large disorder may lead to the many-body localized phase that is characterized by the breakdown of thermalization~\cite{nandkishoreManyBodyLocalizationThermalization2015,Abanin18}. The many-body localized phase can be viewed as the interacting cousin of Anderson insulator~\cite{andersonAbsenceDiffusion}. However, the presence of interactions leads to qualitatively different properties, in particular allowing long-distance entanglement spreading in the many-body localized phase. In addition, by increasing interactions and/or decreasing disorder one can tune the transition between many-body localized and thermalizing phases~\cite{pal10} .

The many-body localization phase transition is a dynamical phase transition that occurs in highly excited eigenstates. This transition separates the thermalizing phase where eigenstates are obeying eigenstate thermalization hypothesis from many-body localized phase characterized by emergent local integrals of motion~\cite{nandkishoreManyBodyLocalizationThermalization2015,Abanin18}. Consequently, one can diagnose the many-body localization phase transition by observing the breakdown of eigenstate thermalization hypothesis in highly excited eigenstates manifested in the scaling of the bipartite entanglement entropy, the bipartite fluctuations of global conserved quantities, participation ratios of many-body eigenstates, and distribution of adjacent energy level spacings~\cite{pal10}.

In this section we  characterize the critical disorder strength~$W_{\rm c}$ as a function of interaction  strength $U$ at which the phase transition between the ergodic and many-body localized phase occurs in the disordered Bose--Hubbard model~(\ref{eq:2_6}). First we introduce the different quantities used to diagnose the phase transition. Afterwards we present the phase diagram, discuss the influence of filling factor, and compare our results to earlier studies.

\subsection{Bipartite entanglement and number uncertainty}
\begin{figure}
    \centering
    \includegraphics[width=1.0\linewidth]{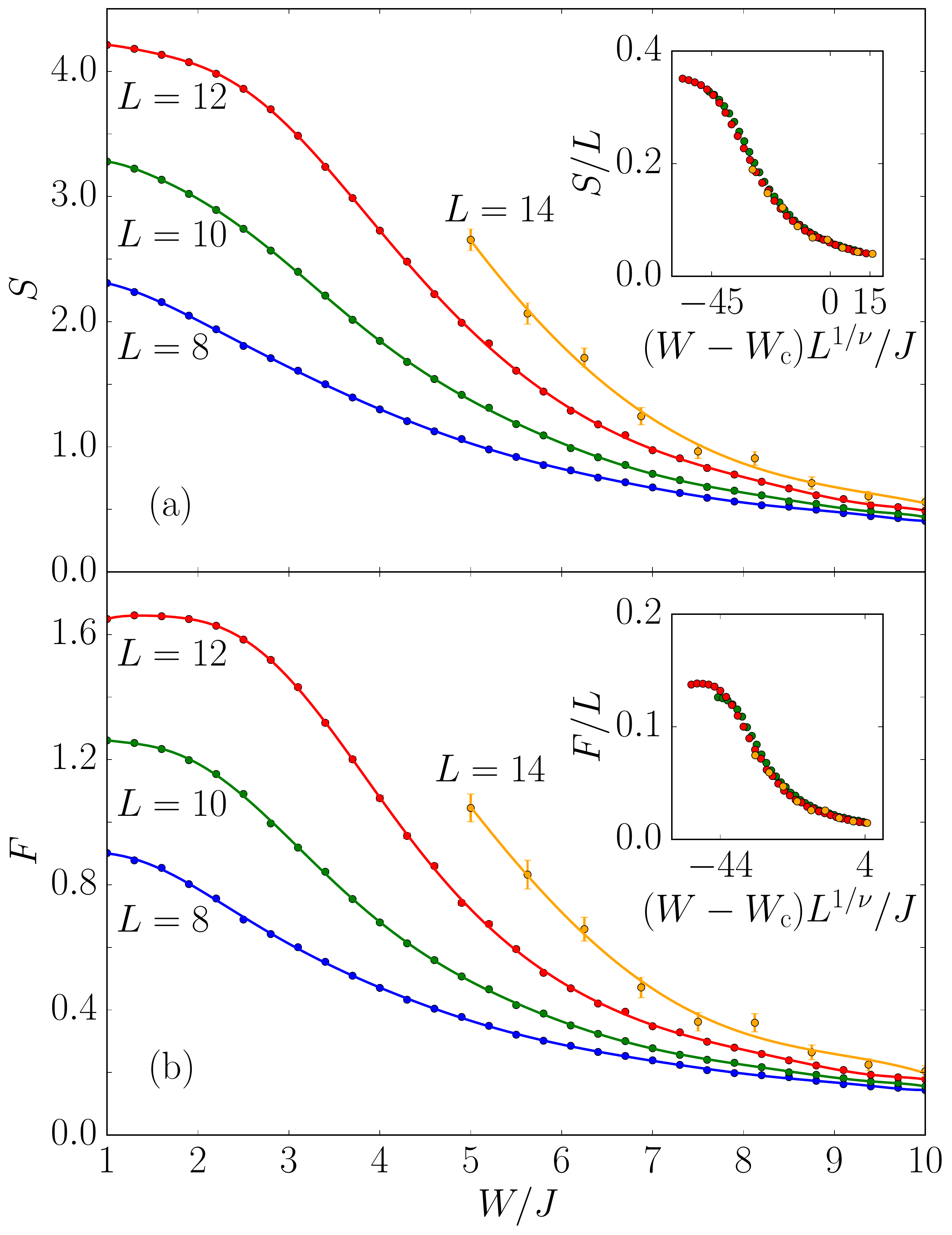}
    \caption{The bipartite entanglement entropy $S$ (a) and the bipartite number uncertainty $F$ (b) as a function of the disorder strength $W$ of uniform disorder distribution for different system sizes $L=8\text{ (blue), } 10\text{ (green), } 12\text{ (red) and } 14\text{ (yellow)}$ with $U/J = 3.5$. The eigenstate, for which the properties are calculated, is the one closest to the maximum density of states. The results are averaged over~\num{4000} disorder realizations, except in $L=14$ we have~\num{264} realizations. Dots denote our data points, and the curves are polynomial fits for visibility. The standard error is denoted with the error bars, and for shorter chains they are smaller than the marker size. In the insets we present the collapsed data from a finite-size scaling analysis with the ansatz~$g[L^{1/\nu}(W - W_{\rm c})]$ without the shortest chain~$L=8$. With this value of $U$, we obtain critical disorder strength~$W_c/J = 8.24$ and scaling~exponent~$\nu = 1.126$ for the bipartite entanglement entropy and for the bipartite number uncertainty~$W_c/J = 9.36$ and~$\nu = 1.24$.}
    \label{fig:3_1}
\end{figure}

The bipartite entanglement entropy between two parts of the system provides a natural quantity to diagnose the transition from ergodic to many-body localized phase. For an arbitrary state~$\ket{\alpha}$ one can construct the corresponding density operator~$\dens = \ket{\alpha}\bra{\alpha}$. We divide the system into two, left and right, partitions. The density operator for the left part of the chain ($A$) is then obtained by tracing out the degrees of freedom of the right part ($B$) from the full density operator,
\begin{equation}\label{eq:3_1}
    \dens_A = \Tr_B\left(\ket{\alpha}\bra{\alpha}\right).
\end{equation}
The entanglement between the subsystems~$A$ and~$B$ is given by the von Neumann entropy of the subsystem density operator, defined as
\begin{equation}\label{eq:3_2}
    S = -\Tr_A\left(\dens_A\ln\dens_A\right),
\end{equation}
and it can be calculated efficiently with Schmidt decomposition~\cite{miszczakSingularValueDecomposition}. In the ergodic phase the entanglement entropy of a typical Hamiltonian eigenstate grows with the size of the subsystem $A$ --- a volume law scaling. In the localized phase, however, the entanglement scales according to an area law since eigenstates can be obtained by a quasi-local unitary acting on a product state~\cite{nandkishoreManyBodyLocalizationThermalization2015,Abanin18}. The transition between these two distinct behaviors provides a tool for diagnosing the phase transition~\cite{luitzManybodyLocalizationEdge2015, singhSignaturesManybodyLocalization2016}.

Even though Hamiltonian~\eqref{eq:2_6} conserves the total number of excitations, the number of excitations within a given half of the system, determined by the operator
\begin{equation}
    \halfop = \sum_{\ell=1}^{L/2}\nop_\ell,
\end{equation}
is not fixed. The fluctuations of $\halfop$ can be used as a characteristic measure between delocalization and localization. The particle number uncertainty is defined through the variance of the half-system particle  number operator $\halfop$,
\begin{equation}\label{eq:3_3}
    F = \braket{\alpha|\halfop^2|\alpha} - \braket{\alpha|\halfop|\alpha}^2,
\end{equation}
and it shows similar behavior as the entanglement entropy in ergodic and localized phases~\cite{luitzManybodyLocalizationEdge2015, singhSignaturesManybodyLocalization2016, songBipartiteFluctuations}.

We have studied how the disorder strength~$W$ and the on-site interaction~$U$ affect the bipartite entanglement entropy $S$ and the bipartite number uncertainty $F$ in different sized systems for equal bipartition of the transmon chain: data for~$U/J=3.5$--value is shown in Fig.~\ref{fig:3_1}. The eigenstate~$\ket{\alpha}$ for which these properties are calculated is selected to be the one closest to the estimated maximum of the density of states for each disorder realization. We observe that with weak disorder both observables scale according to the volume law, but as the disorder increases the scaling turns to the area law, which signals presence of a phase transition from ergodic to localized phase. For large on-site interaction~$U$ the eigenstates form the miniband structure, as discussed in Sec.~\ref{sec:transarray}. In such a situation the density of states has several local maxima, and the selection of the eigenstates is no longer clear. However, the miniband structure rapidly vanishes with increasing~$W$ for the studied values of~$U$, and therefore it does not affect the results. 

\subsection{Energy level statistics}
\begin{figure}
    \centering
    \includegraphics[width=1.0\linewidth]{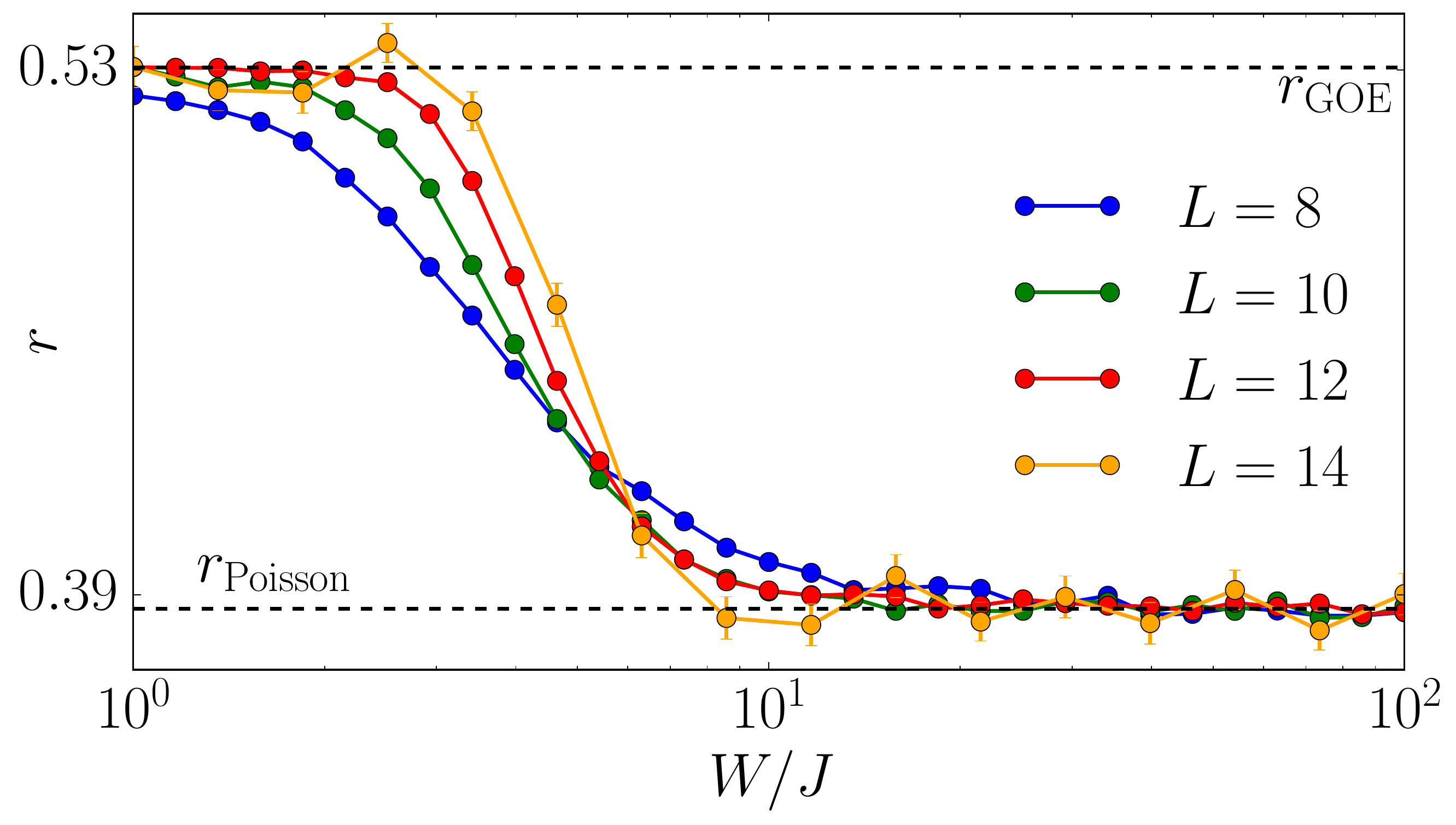}
    \caption{Average adjacent gap ratio~$r$ as a function of the disorder strength $W$ for different sized systems with anharmonicity~$U/J = 3.5$. The reference values for Gaussian orthogonal ensemble~($r_{\rm GOE}$) and Poissonian distribution~($r_{\rm Poisson})$ are denoted by black horizontal lines. The results are averaged over~4000 realizations, except for~$L=14$ which is averaged over~264 realizations. The eigenvalues are located at the maximum density of states, and the gap ratio is calculated over~16 adjacent eigenstates from each disorder realization. The error bars denote the standard error and are in general smaller than the marker size.}
    \label{fig:3_3}
\end{figure}
 The bipartite entanglement entropy and bipartite number uncertainty are properties of the Hamiltonian eigenstates, but the distinction between ergodic and localized phases is visible also in the distribution of the energy eigenvalues. A widely used tool for measuring it is the energy level spacing distribution. In the ergodic phase the eigenvalues are distributed according to the Gaussian orthogonal ensemble, while in the localized phase they are uncorrelated and obey Poissonian statistics~\cite{ChaosReview}. One often considers the adjacent gap ratio~\cite{luitzManybodyLocalizationEdge2015, mondainiManybodyLocalizationThermalization2015, oganesyan07, pal10}
\begin{equation}\label{eq:3_4}
    r^{(n)} = \frac{\min\left[\delta^{(n)}, \delta^{(n+1)}\right]}
    {\max\left[\delta^{(n)}, \delta^{(n+1)}\right]},
\end{equation}
where $\delta^{(n)} = E_n - E_{n-1} > 0$ is the energy difference between a pair of adjacent levels. In the ergodic phase the average adjacent gap ratio over $n$ is~$r_{\rm GOE}\approx 0.536$, and in the localized phase~$r_{\rm Poisson}\approx 0.386$~\cite{atas13}. We have also studied the distribution of the adjacent gap ratios as a function of disorder strength around the maximum of the density of states. At weak disorder the average gap ratio is consistent with the Gaussian orthogonal ensemble~\cite{atas13}, whereas for strong disorder energy levels become essentially uncorrelated and the average gap ratio tends to $r_{\rm Poisson}$, as shown in Fig.~\ref{fig:3_3}. An estimate for the transition point can be obtained from the point at which the lines of different sized systems cross. However, in order to obtain convergent results for the transition point one has to calculate a large number of eigenvalues from each realization, which makes the adjacent gap-ratio computationally much more expensive than the bipartite entanglement and number uncertainty for longer chains. For this reason we have not used the average adjacent gap ratio in further studies of the phase transition.

\begin{figure}
    \centering
    \includegraphics[width=1.0\linewidth]{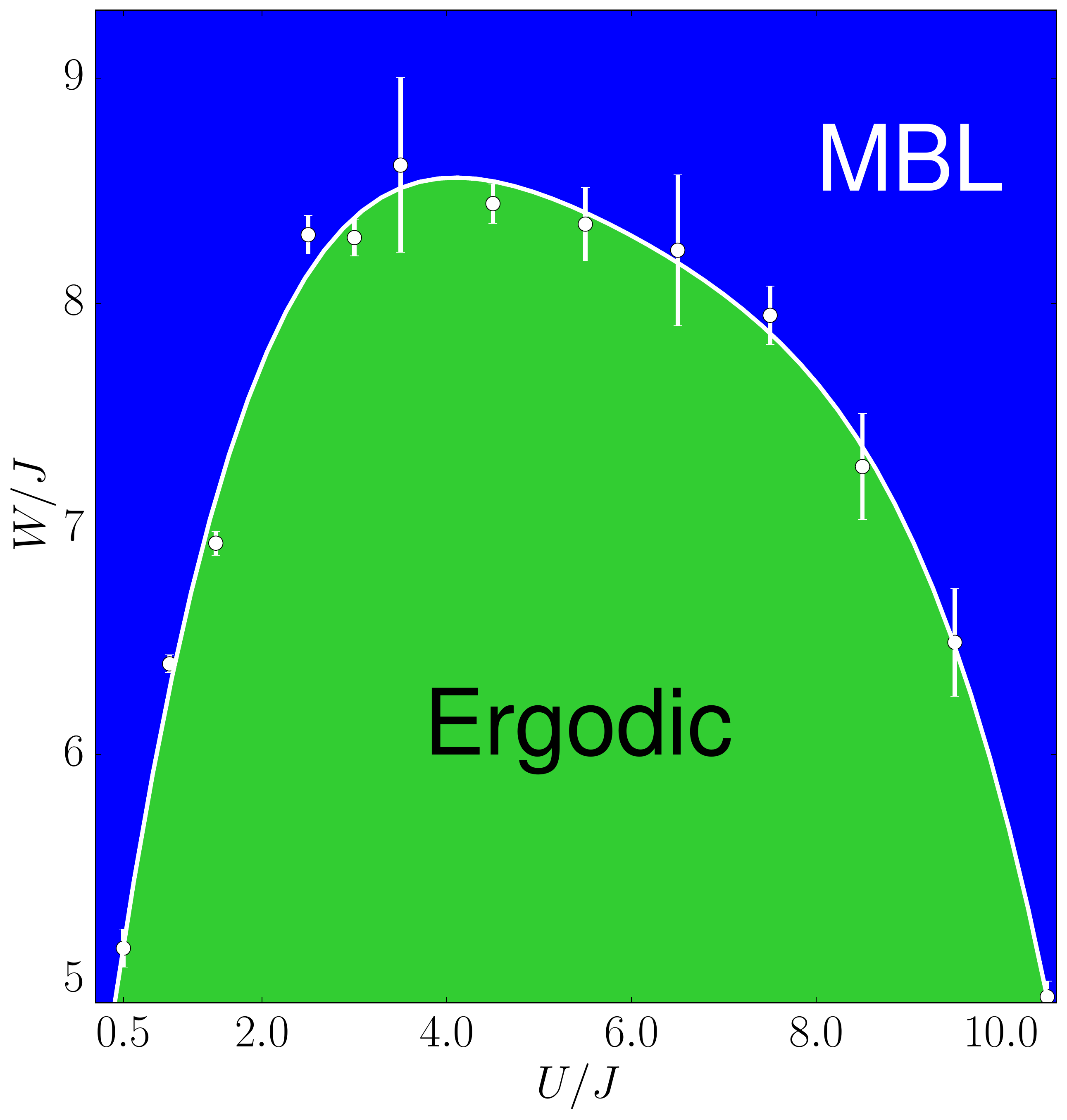}
    \caption{Phase diagram  for the half-filled disordered Bose--Hubbard Hamiltonian~\eqref{eq:2_6} as a function of the on-site interaction $U$ and the disorder strength $W$. The transition point is estimated using finite-size scaling analysis for different sized systems $L=10,12,14$ for the bipartite entanglement entropy and the bipartite number uncertainty, shown in Fig.~\ref{fig:3_1}. The disorder is included only in the on-site energy $\omega_\ell$ and it is drawn from the uniform distribution~$[-W, W]$. The data points are obtained as an average of the transition points given by the finite-size scaling analysis on the bipartite entanglement entropy and the bipartite number uncertainty and the error bars show the deviation between these two values. The white curve is a polynomial fit that improves visual clarity.}
    \label{fig:3_2}
\end{figure}

\subsection{Phase diagram}
An estimate for the critical disorder strength~$W_{\rm c}$ can be obtained from the data in Fig.~\ref{fig:3_1} by scaling the curves with corresponding chain lengths and determining the disorder strength at which the curves cross. Another possibility, the one that we use, is the finite-size scaling collapse using ansatz~$g[L^{1/\nu}(W - W_{\rm c})]$, which collapses the data to a single curve~\cite{luitzManybodyLocalizationEdge2015, kudoFiniteSize1018}(see insets of Fig.~\ref{fig:3_1}). We obtain the critical disorder strength as a function of the anharmonicity~$U$ for both the bipartite entanglement entropy  and the bipartite number uncertainty, and use their average as an estimation for the transition point. The phase-diagram shown in Fig.~\ref{fig:3_2} is constructed from this data. The overall shape of the phase diagram is similar to the corresponding fermionic system studied in Ref.~\onlinecite{mondainiManybodyLocalizationThermalization2015}. However, the maximum critical disorder strength is reached at much weaker on-site interactions in the bosonic system compared to the fermionic case. We attribute this behavior to the fact that the on-site interaction is effectively much stronger in the bosonic case since the number of excitations per site is not limited to one. 

From the point of view of experimentally realizable superconducting transmon circuits, the attainable parameter regime (see Table~\ref{tab:2_1}) is roughly $U/J=2-30$ and $W/J=0.1-200$ yielding that the phase transition occurs within the experimentally realizable parameter range. Furthermore, based on our additional calculations (not shown here) with experimentally relevant parameters, the additional perturbations by the higher-order anharmonicity in Eq.~\eqref{eq:2_3} and the next-nearest neighbor hopping in Eq.~\eqref{eq:2_2} do not change the situation and the phase diagram remains intact within the original error bars. However, if the next-nearest neighbor hopping~$J_2$ is made strong enough~$J_2\gtrsim J/5$, the transition point shifts towards larger disorder strengths. This can be qualitatively understood as an increase in kinetic energy, promoting delocalization~\cite{sierantManybodyLocalizationPresence2019}. The higher order on-site interaction~$U_2$, on the other hand, effectively weakens the anharmonicity due to opposite sign. With sufficiently large values~$U_2\gtrsim U/3$, the higher-order anharmonicity shifts the transition points towards larger~$U$. Finally, in systems where the disorder is included also in the hopping and on-site interaction terms the phase transition occurs at slightly weaker disorder strengths since the system is more disordered.

We have studied the phase transition also for systems with the transmon-disorder of Eq.~\eqref{eq:2_5} and observed that the critical disorder strength corresponds to a somewhat larger~($\lesssim J$) disorder strength than with the uniform disorder. This happens because disorder distributions with larger standard deviations are more effective at localizing the system. Indeed, in the studied disorder strength range the uniform distribution has larger standard deviation than the transmon disorder (magnetic field values between~\num{0.01} and~\num{0.3} in Fig.~\ref{fig:2_0}). With large anharmonicities the on-site interaction starts to dominate, and the critical disorder strength is within error bars the same for both disorder distributions.

Increase of the filling factor also increases the critical disorder strength. We attribute this to the bosonic enhancement of tunneling for multiply occupied sites which makes the system more robust against localization. We have performed simulations for small systems ($L<10$) with unit filling and confirmed this behavior. The numerical simulations with larger fillings are limited to smaller system sizes than in the half-filled case due to larger total Hilbert spaces, given by Eq.~\eqref{eq:2_states}. Because there is no upper limit for the total number of excitations in bosonic systems, an open question remains how critical disorder strength behaves at much larger fillings.

The phase transition was studied experimentally in Ref.~\onlinecite{roushanSpectroscopicSignaturesLocalization2017}, where a chain of nine transmons with two excitations in total was studied through adjacent energy gap and participation ratios. They used quasi-periodic potential~$\omega_\ell = \Delta\cos(2\pi\beta\ell)$, where~$\Delta$ is the disorder strength and~$\beta$ is an irrational number. Notice that the quasi-periodic potential with the disorder strength $\Delta$ has standard deviation $\Delta/\sqrt{2}$, i.e. using standard deviation as a disorder measure the quasi-periodic potential is more random than the uniform disorder with disorder amplitude $W=\Delta$ and standard deviation $\Delta/\sqrt{3}$. For~$U/J = 3.5$ Roushan \emph{et al.}~in Ref.~\onlinecite{roushanSpectroscopicSignaturesLocalization2017} found a phase transition at~$\Delta/J\approx2$, which corresponds to~$W/J\approx 2.5$ for the uniform potential when the mapping is done through equal standard deviations between the two distributions. Such a small disorder strength compared to our phase diagram is explained partly by the small filling factor and partly by the more disordered potential.

\section{Dynamics} \label{sec:dyn}
All the eigenstate measures of localization that we have presented so far (eigenstate bipartite entanglement entropy, bipartite number uncertainty and average adjacent gap ratio in Figs.~\ref{fig:3_1}-\ref{fig:3_3}) are shared between the phases of Anderson localization and many-body localization. The distinction between the two phases of localization as well as between the localized and ergodic phases can be observed in several dynamical properties, some of which are available also for experiments on superconducting circuits. In this paper we consider only unitary dynamics. We utilize exact eigendecomposition only for the shortest system $L=8$ and for the larger systems $L=10,12,14$ the unitary time evolution is calculated through Krylov subspace methods, see App.~\ref{app:time} for details. Furthermore, we simulate the time evolution for a long chain of $L=40$ transmons to confirm that the results are properly saturated in the system size and exhibit no boundary effects. The long chain simulations are performed using time evolving block decimation scheme for matrix product states, detailed in App.~\ref{app:timeMPS}.

Dissipation and decoherence are always present in superconducting circuits, and they are expected to eventually destroy the many-body localized phase~\cite{medvedyeva16, levi16, luschen17}. However, due to long relaxation and coherence times of modern transmons~($T_1 \approx $~\SI{60}{\micro\second} and~$T_2 \approx $~\SI{20}{\micro\second})~\cite{Paik11}, the system dynamics remain unitary to relatively long times, about~\SI{10}{\micro\second}. Based on our simulations this is sufficient for observing the distinct behaviors of all three phases. 

\subsection{Dynamics of bipartite entanglement entropy}

\begin{figure*}
    \centering
    \includegraphics[width=1.0\linewidth]{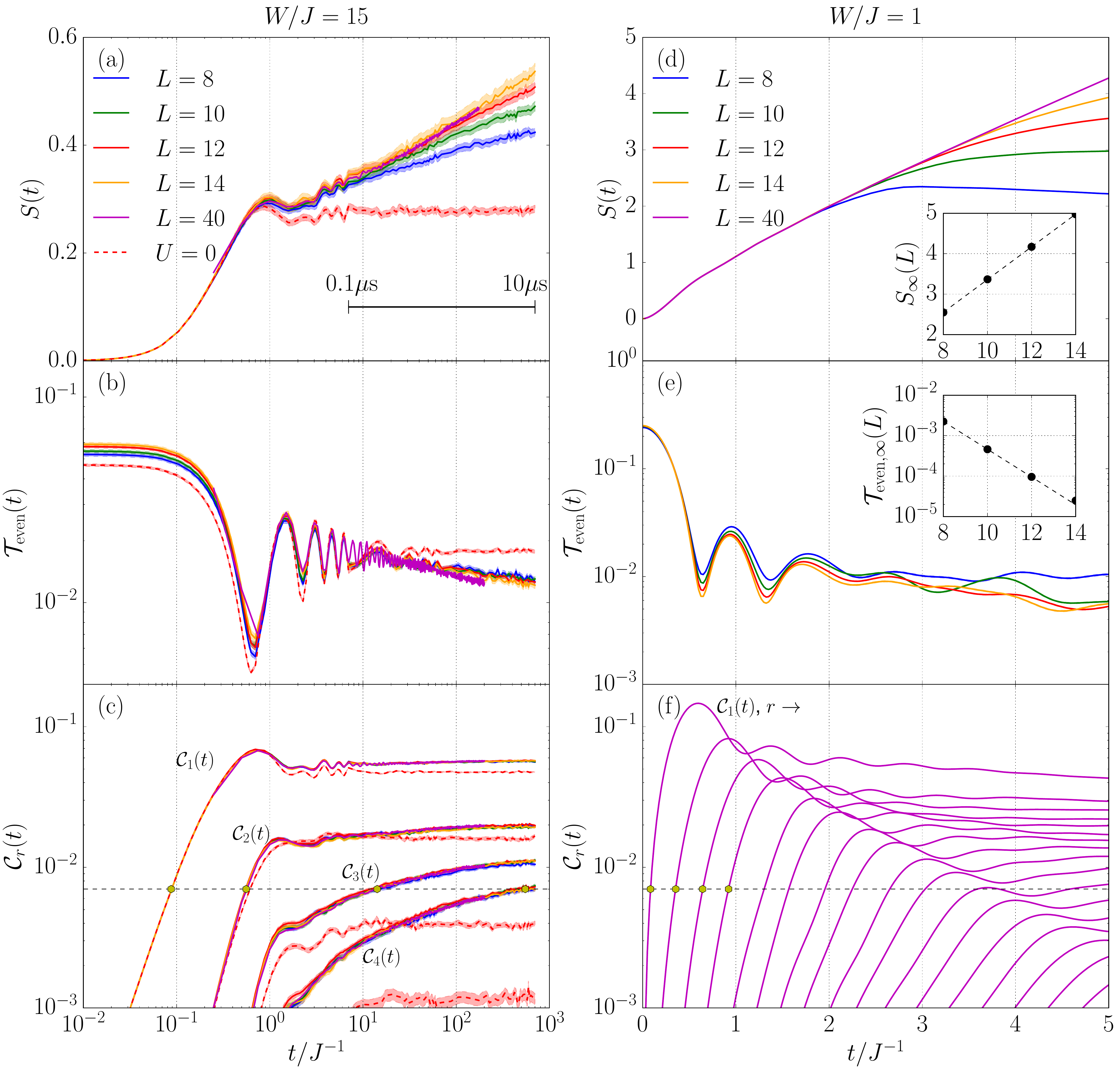}
    \caption{The bipartite entanglement entropy~$S(t)$ [(a) and (d)], the temporal number fluctuation~$\tempflu_{\rm even}(t)$ of even sites [(b) and (e)], and the two-site number correlations~$\mathcal{C}_r(t)$ [(c) and (f)] as a function of time~$t$. The dynamical probes are calculated for interacting system with $U/J = 3.5$ (solid lines) for system sizes $L=8 \text{ (blue), }10\text{ (green), }12\text{ (red), }14\text{ (yellow)} \text{ and }40\text{ (purple)}$, and for the non-interaction situation $U/J = 0$ with $L=12$ sites (dashed red line). The disorder distribution is uniform, and panels (a)--(c) correspond to strong disorder $W/J=15$ and panels~(d)--(f) to weak disorder $W/J=1$. Results for system sizes $L\le12$ are averaged over~2000, for~$L=14$ over~1000 and for $L=40$ over 500 disorder realizations. For $L=40$ additional truncation of on-site Hilbert space was applied with maximal occupancy limited to $n_{\rm max}= 4$ for $W/J=1$ and $n_{\rm max}= 3$ for $W/J=15$, which is justified by the dilute initial state and rather short evolution times. Shaded regions depict the standard errors of the disorder averages. In~(c) and~(f) the yellow dots depict the points at which the corresponding correlations reach the predetermined reference value (dashed horizontal line). The insets in (d) and (e) describe the equilibrium values of the corresponding observables as a function of system size. Note also the different time scales in left and right columns.}
    \label{fig:4_1}
\end{figure*}

Let us start by considering the quench dynamics under the disordered Bose--Hubbard Hamiltonian of Eq.~\eqref{eq:2_6}. We perform the time evolution of the non-entangled initial state $\ket{\psi_0}$ and measure the dynamics of the entanglement entropy~$S(t)$. Our choice for the initial state is a N\'eel-type of state~$\ket{101010\dots}$ studied also in experimental setups~\cite{Xu17}. The dynamics of the bipartite entanglement entropy $S(t)$ for different sized systems is shown in Fig.~\ref{fig:4_1}(a),(d). Our main interest is in the differences of the dynamical behavior in the many-body localized, Anderson localized and ergodic phases.

At strong disorder in Fig.~\ref{fig:4_1}(a), the early dynamics of the bipartite entanglement entropy $S(t)$  for the interacting (many-body localized) and non-interacting (Anderson localized) systems are similar, that is, they both initially exhibit similar rapid growth of entanglement. However, for the interacting system the information spreading does not stop after reaching the length scales of the localization length but instead continues at much smaller rate than initially. This logarithmic growth of entanglement at long times is caused by the interaction-induced dephasing~\cite{singhSignaturesManybodyLocalization2016, znidaricManybodyLocalization, bardarsonUnboundedGrowthEntanglement, nanduriEntanglementSpreading, serbynSlowGrowthEntanglement} present only in the many-body localized phase. The interacting system eventually equilibrates, but not to the canonical ensemble~\cite{Alet18, goihlExperimentallyAccessibleWitnesses2016}. Thus, in the many-body localized phase there exists two distinct regions in time evolution. During the first one the excitation quanta explore the system within the localization length. After this they start to slowly dephase with other particles further apart, which leads to the described long-time behavior. 

At weak disorder in Fig.~\ref{fig:4_1}(d), information rapidly spreads and eventually reaches thermal equilibrium as expected in the ergodic phase~\cite{ChaosReview}. The entanglement entropy~$S(t)$ has a ballistic growth and saturates to a value that obeys the volume law~\cite{kimBallisiticSpreadingEntanglement} [inset of Fig.~\ref{fig:4_1}(d)]. After the saturation and thermal equilibration, local observables are determined by canonical ensemble~\cite{luitzErgodicSideManybody, Alet18}. In the Anderson localized phase at $U=0$ [dashed line in Fig.~\ref{fig:4_1}(a)] the dynamics is constrained within the localization length and the entanglement entropy saturates to a much smaller value than in the ergodic phase. This saturation value depends on the localization length and thus behaves according to the area law (not shown here). Anderson localized systems never reach thermal equilibrium within the whole system~\cite{goihlExperimentallyAccessibleWitnesses2016}.
 
\subsection{Dynamics of on-site number fluctuations}
The entanglement entropy provides a good measure for identifying the many-body localized phase, and recent experiments with optical lattices~\cite{lukinProbingEntanglementManybody2019} and trapped ions~\cite{brydgesProbingRenyiEntropy} have demonstrated that it can also be measured without mapping the full density matrix. We expect that similar schemes can be extended also to superconducting circuits. However, it is still beneficial to study simple observables that require measurements of only a few local expectation values and show the distinction between the three phases. Such observables are more accessible in experiments. One possibility is to study temporal fluctuations of local observables~\cite{serbynQuantumQuenchesManybody2014, detomasiSolvingEfficientlyDynamics2018}. We consider here dynamics of fluctuations for the number operator~$\nop_\ell$ of the $\ell$-th site, defined as  
\begin{equation}\label{eq:4_1}
    \tempflu_\ell(t) = \left\langle\left[\braket{\nop_\ell(t)} - \bar n_\ell\right]^2\right\rangle_{\rm d},
\end{equation}
where $\langle\cdot\rangle_{\rm d}$ denotes average over disorder realizations. The fluctuations are measured with respect to a steady-state, equilibrium value defined here as a longtime average
\begin{equation}\label{eq:4_2}
    \bar n_\ell = \lim_{t\to\infty}\frac{1}{t}\int_0^t\braket{\nop_\ell(\tau)}d\tau.
\end{equation}
For additional convergence we have averaged the temporal fluctuations over even and odd sites separately, and denoted these averages as $\tempflu_{\rm even}(t)$ and $\tempflu_{\rm odd}(t)$, but the results are similar also for individual sites. Fluctuations averaged over even sites are shown at strong disorder in~Fig.~\ref{fig:4_1}(b) and at weak disorder in~Fig.~\ref{fig:4_1}(e). At weak disorder, only results from the exact diagonalization are shown, since the time evolving block decimation is restricted to short times and thus the long-time average is not accessible. Experimental advantage of this measure is that it requires only measurement of a single site or few sites and it can be achieved in high-precision with superconducting circuits through dispersive readout that naturally measures transmon occupation $\braket{\nop_\ell}$~\cite{Krantz19}.

Monitoring temporal fluctuations of a local operator can be seen as a measure for attainable volume that an initial excitation can explore. In Anderson localized systems, temporal fluctuations never vanish~\cite{detomasiSolvingEfficientlyDynamics2018} since the effective volume is strictly limited by the localization length, seen in Fig.~\ref{fig:4_1}(b) (dashed line). In the many-body localized phase, after transients, the fluctuations decay as the power-law [Fig.~\ref{fig:4_1}(b)]
\begin{equation}
    \tempflu_\ell(t) \propto t^{-b},
\end{equation}
being a signal that after the initial build-up of the localization wave function the effective volume is slowly expanding~\cite{serbynQuantumQuenchesManybody2014, detomasiSolvingEfficientlyDynamics2018}. Curiously, there is a clear distinction at later times between the power law like decay in the many-body localized phase and the saturation in the Anderson localized phase, but this distinction becomes visible roughly one order of magnitude later than in the bipartite entanglement entropy. In thermal phase [Fig.~\ref{fig:4_1}(e)] we observe a rapid decay of the fluctuations until saturation to a value that depends on the system size as~$e^{-aL}$, where~$a$ is some positive constant [inset of Fig.~\ref{fig:4_1}(e)].

\subsection{Dynamics of two-site correlations}
Finally, we study the propagation of information using two-site number correlations~\cite{detomasiSolvingEfficientlyDynamics2018, goihlExperimentallyAccessibleWitnesses2016}. For two sites separated by distance~$r$ the correlations are defined as
\begin{equation}\label{eq:4_3}
    \cor_{\ell,r}(t) = \big\langle\left|\braket{\nop_\ell(t)\nop_{\ell+r}(t)} -     \braket{\nop_\ell(t)}\braket{\nop_{\ell+r}(t)}\right|\big\rangle_{\rm d}.
\end{equation}
In order to improve convergence we average over each pair with fixed separation~$r$ to obtain the distance dependent correlation~$\cor_r(t)$,
\begin{equation}
    \cor_r(t) = \frac{1}{L-r}\sum_\ell\cor_{\ell, r}(t),
\end{equation}
where $L-r$ is the number of $r$ separated pairs in chain of $L$ sites. The conclusions are similar also for individual pairs.  Experimentally this might be more challenging to measure than the temporal fluctuations due to the two-site correlation.

The time evolution for the four nearest correlations~$\cor_1(t)-\cor_4(t)$ is shown in Fig.~\ref{fig:4_1}(c) for strong disorder for different system sizes. In Fig.~\ref{fig:4_1}(f) we display the weak disorder case for~$L=40$ and additional longer range correlations. In both figures we have also displayed a reference value (dashed horizontal line) for monitoring the development of nearest correlations~$\cor_1(t)-\cor_4(t)$ in different phases. In thermal phase the correlations first develop between adjacent sites and spread to more distant pairs at constant velocity, i.e. the peaks of the correlations are linearly spaced in time. Similarly, the correlations reach the reference value at equidistant times. The height of the peaks decays exponentially with the distance~$r$, which implies exploration of finite fraction of Hilbert space up to Lieb-Robinson bounds~\cite{lieb1972finite}. After rapid initial dynamics, the correlations saturate to values inversely proportional to the system volume. The equilibrium values are all of the same magnitude.

In the localized systems the correlations first increase rapidly. After the localization length is reached, the correlations in the Anderson localized system saturate to a value inversely proportional to the separation of the pairs. Saturated values obey the area law (not shown), and there is a several orders of magnitude difference between different distances. In the many-body localized system the correlations instead continue to grow at much smaller rates than initially. Because of this slow growth, the reference value for correlations at different distances is reached at logarithmically spaced times, revealing a logarithmic light cone~\cite{detomasiSolvingEfficientlyDynamics2018}. On the other hand, in the Anderson localized systems the correlations never reach the reference value due to the saturation.

In summary, we conclude that the interacting and strongly disordered system resulting many-body localization is clearly distinguishable from that of non-interacting and ergodic phases by the presented dynamical probes shown in Fig.~\ref{fig:4_1}. Importantly, this dynamical distinction between the phases occurs at experimentally feasible time-scales \SIrange{0.1}{10}{\micro\second} [horizontal bar in Fig.~\ref{fig:4_1}(a)] set by the decoherence and dissipation rates of the modern superconducting ciruits. 

\section{Conclusions}~\label{sec:conc}
In this work we have numerically studied the many-body localization phase transition in the attractive Bose--Hubbard Hamiltonian using the methods of exact diagonalization as well as matrix product state dynamics. Such systems can be experimentally realized with arrays of superconducting circuits, and our purpose was to produce results that could be verified experimentally with currently available technology. The distinct features of many-body localization are visible in systems with a minimum of eight transmons, as shown in Fig.~\ref{fig:4_1}, although similar signatures can be observed already with six sites. 

The bipartite entanglement entropy, bipartite number uncertainty and adjacent gap ratio of the Hamiltonian eigenpairs exhibit ergodic behavior at weak and localized behavior at strong disorder. Using finite size scaling analysis we have obtained an estimate for the critical disorder strength as a function of the transmon anharmonicity and constructed the ergodic--many-body localized phase diagram for the attractive Bose--Hubbard Hamiltonian~(Fig.~\ref{fig:3_2}). The phase transition occurs at experimentally feasible parameters and it is robust against higher order on-site interactions and longer range tunneling. The eigenstates were taken at the maximum density of states, which, due to the anharmonicity and bosonic nature of the system, is heavily dependent on the Hamiltonian parameters as well as on the disorder realization. Thus, in order to study comparable eigenstates we had to estimate the density of states for each realization. For smaller systems this was done with the $LDL$ decomposition method that resorts to Sylvester's law of inertia. For larger systems an approximation with the stochastic Chebyshev expansion was used.

Distinction between the many-body localization and Anderson localization of non-interacting systems can be observed in the dynamics. We have simulated unitary quench dynamics and studied the time evolution of entanglement entropy, temporal number fluctuations and two-site number operator correlations. These observables feature distinct behavior in ergodic, many-body localized and Anderson localized phases. In the many-body localized phase the entanglement displays logarithmic growth at long times, the temporal fluctuations decay according to a power law and the correlations spread logarithmically to more distant sites. This behavior becomes visible at experimentally relevant time scales, and thus we suggest that the temporal fluctuations and correlation functions are suitable dynamical observables for the experimental studies of the many-body localization in systems of superconducting circuits. We believe that the results presented in this paper will increase the attention and lead to focused experimental studies of many-body localization in systems of superconducting circuits. 

The localization phenomenon ideally occurs in closed systems but it is known to survive for intermediate times in weakly open systems~\cite{Johri15, Fischer16, luschen17}. Dissipation and decoherence will have primarily different roles in the many-body localization of transmon arrays. Dissipation removes energy and excitations from the system, eventually bringing it into a dilute non-interacting phase, whereas
 decoherence destroys localization by destroying the phase coherence. The models for dissipation and decoherence of superconducting transmons are well known and characterized~\cite{Devoret13}, which makes it an excellent basis for studying open quantum system effects on many-body localization both experimentally and theoretically.  Specific research questions are, for example, what time scales localization will survive under dissipation and decoherence, and  how these times depend on the filling factor, interaction and disorder strengths, or how does continuous monitoring affect localization through measurement back action. We leave addressing these questions for a future work.

\section*{Acknowledgements}
Special thanks are addressed to R.~T.~Brierley and S.~M.~Girvin for discussions and contributions in the early phase of the project. We are also grateful for fruitful discussion with Bryan K.~Clark, Claudia De Grandi, Gerhard Kirchmair, Zaki~Leghtas, Olli Mansikkam\"aki, Stefan Oleschko, Iivari~Pietik\"ainen, Erkki Thuneberg, Sasu Tuohino, Jani~Tuorila, and Xiongjie Yu.  This research was financially supported by the Alfred Kordelin Foundation, the Emil Aaltonen Foundation, and the Academy of Finland under grant nos.~316619~and~320086. We also wish to acknowledge CSC--IT~Center~for~Science,~Finland, and Finnish Grid and Cloud Infrastructure (persistent identifier urn:nbn:fi:research-infras-2016072533) for computational resources. 

\appendix 

\section{Exact diagonalization} \label{app:ex}
We are interested in the properties of the eigenstates and eigenvalues of the Hamiltonians of Eqs.~\eqref{eq:2_1} and~\eqref{eq:2_6}. Due to the large dimensions and the need for several disorder realizations, the efficient full diagonalization is limited to small systems~$L\le10$. However, we do not need the full spectrum, but only eigenpairs close to some specified target energy. The target-specified eigenpair can be obtained efficiently with the shift-and-invert method, where one considers an eigenvalue problem 
\begin{equation}
    \mat H\vec u = \lambda\vec u, 
\end{equation}
where the eigenvector~$\vec u$ is such that the eigenvalue is close to the target, $\lambda\approx \sigma$. To obtain this eigenpair efficiently, we can make a spectral shift and consider the matrix~$\left(\mat H - \sigma\mat I\right)^{-1}$. The largest eigenvalue of the matrix $\left(\mat H - \sigma\mat I\right)^{-1}$ corresponds to the eigenvalue of the matrix~$\mat H$ closest to the target~$\sigma$. 

Thus, we obtain a new eigenvalue problem
\begin{equation}
    \left(\mat H - \sigma\mat I\right)^{-1}\vec u = \frac{1}{\lambda - \sigma}\vec u,
\end{equation}
The eigenvector~$\vec u$ can then be obtained with the power iteration method where one repeatedly applies the matrix~$\left(\mat H - \sigma\mat I\right)^{-1}$ to an initial random vector~$\vec u_0$ and normalizes the result:
\begin{equation}
    \vec u_{k+1} = \frac{\left(\mat H - \sigma\mat I\right)^{-1}\vec u_{k}}
    {||\left(\mat H - \sigma\mat I\right)^{-1}\vec u_{k}||}.
\end{equation}
After a suitable number of iterations one obtains the eigenvector of the original matrix with the eigenvalue closest to the target. Because inverting a large matrix is a challenging operation, it is customary to convert the inverted matrix-vector multiplication to a system of linear equations subsequently solved with the $LU$ decomposition~\cite{luitzManybodyLocalizationEdge2015, pietracaprinaShiftinvertDiagonalizationLarge2018a, luitzErgodicSideManybody}, where~$\mat L$ and $\mat U$ are lower and upper triangular matrices. We have~$\left(\mat H - \sigma\mat I\right)\vec w = \mat L\mat U\vec w = \vec u$, which results in~$\vec w = \left(\mat H - \sigma\mat I\right)^{-1}\vec u$. We use the shift-and-invert method provided by Spectra library~\cite{spectraweb} built on top of Eigen library~\cite{eigenweb}. This algorithm also transforms the matrix inversion to a system of linear equations and has the advantage that it can give arbitrary number of states closest to the target.

\section{Efficient estimation of the density of states} \label{app:LDL}
In order to use the shift-and-invert method, one needs a target eigenvalue. Since we are interested in the eigenstate closest to the maximum density of states, we need to know the location of the maximum of the density of states. If the full eigen decomposition is possible, the density of states is obtained as a trivial side product. However, more efficient way of estimating the density of states is to apply Sylvester's law of inertia~\cite{napoliEfficientEstimationEigenvalue2016} which gives the number of eigenvalues of a matrix $\mat{H}$ within an arbitrary interval~$[\varepsilon_i, \varepsilon_{i+1}]$. 
 
In our case, $\mat H$ is real, symmetric and non-singular matrix with a well-defined $LDL$ decomposition
\begin{equation}
    \mat H = \mat L\mat D\mat L^T, 
\end{equation}
where $\mat L$ is a lower unit triangular matrix and~$\mat D$ is a diagonal matrix. Sylvester's law of inertia states that the number of positive diagonal entries of the matrix~$\mat D$ is equal to the number of positive eigenvalues of the matrix~$\mat H$. One can then construct the shifted matrices $\mat H - \varepsilon_i\mat I$ and~$\mat H - \varepsilon_{i+1}\mat I$, whose $LDL$ decompositions give the number of eigenvalues of~$\mat H$ that are larger than~$\varepsilon_i$ and~$\varepsilon_{i+1}$, respectively. Difference of these numbers gives the exact number of eigenvalues within the interval~$[\varepsilon_i, \varepsilon_{i+1}]$, which can be used to construct the density of states. The algorithm that we use to perform the $LDL$ decomposition is included in Eigen library~\cite{eigenweb}.

Disadvantage of applying Sylvester's law of inertia is the need to perform several full $LDL$ decompositions in order to determine accurate the density of states. This limits the applicability of the method only to moderately small system sizes,~$L \lesssim 10$. For larger systems one has to resort to approximations. Our choice is a method that relies on Chebyshev series~\cite{napoliEfficientEstimationEigenvalue2016}. In this method, one considers a projection operator 
\begin{equation}\label{eq:B_proj}
    \hat P_{\varepsilon_i, \varepsilon_{i+1}}= \sum_{\lambda_j \in [\varepsilon_i, \varepsilon_{i+1}]} \ket{u_j} \bra{u_j}
\end{equation}
constructed from eigenvectors $\ket{u_j}$ whose eigenvalues $\lambda_j$ are within the interval~$[\varepsilon_i, \varepsilon_{i+1}]$. The trace of this operator then gives exactly the number of eigenvalues within this interval of interest. Since the eigenvectors are not known a priori, one has to approximate both the operator and its trace. Our discussion here follows closely to that given in Ref.~\onlinecite{napoliEfficientEstimationEigenvalue2016}.

\begin{figure}
    \centering
    \includegraphics[width=1.0\linewidth]{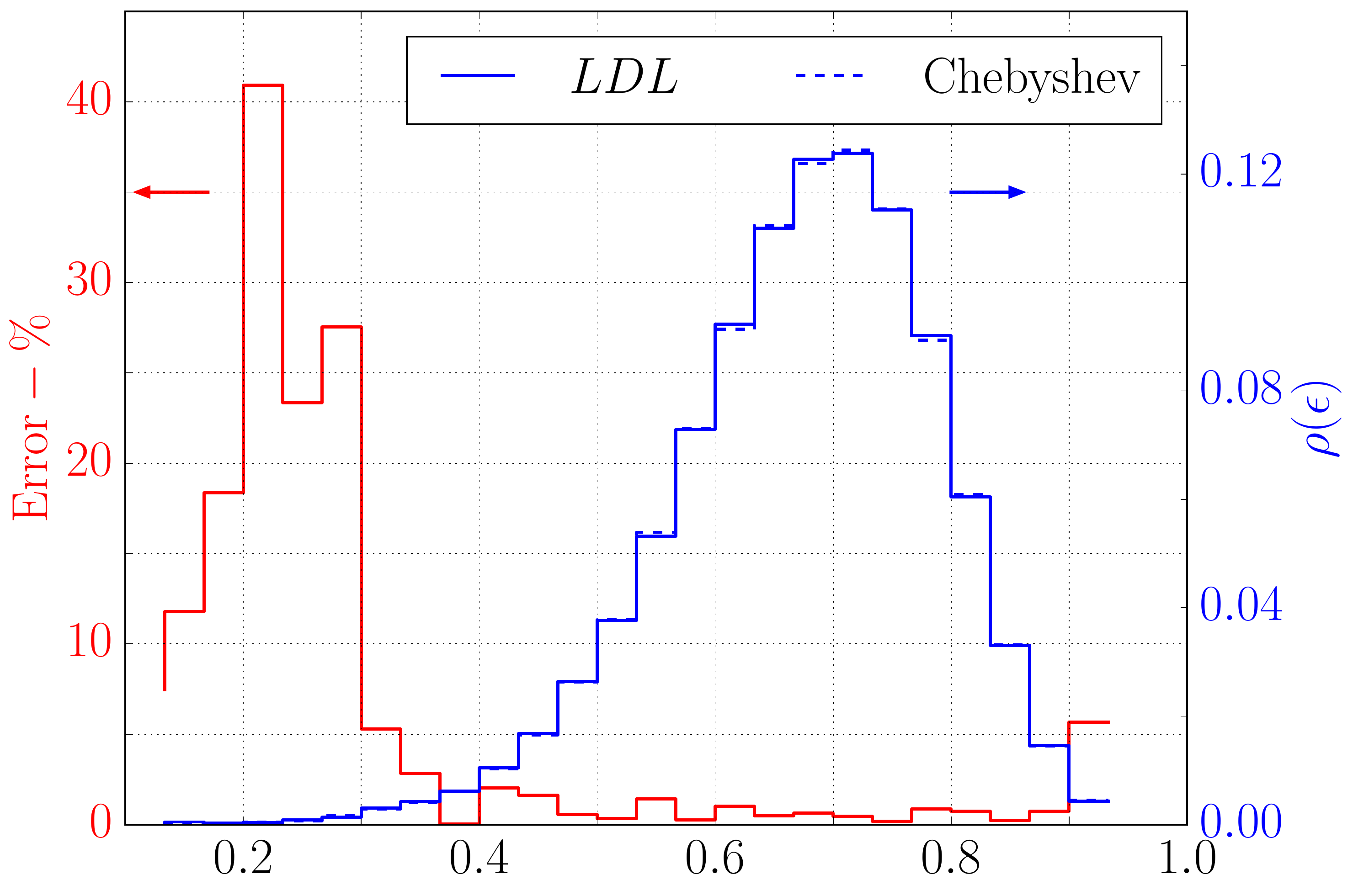}
    \caption{Comparison of the density of states histograms for system with~$L=12$ using $LDL$ decomposition (solid blue) and Chebyshev expansion (dashed blue). Relative error percentage is shown as the red histogram and it tells how much the results obtained with the Chebyshev expansion differs from those obtained with the $LDL$ method. Chebyshev method is the most accurate in intervals with large number of states. Around the maximum density of states the error is~$\sim 1\%$. In the Chebyshev expansion we use~\num{50} terms and the stochastic trace is obtained with~\num{30} random vectors.}
    \label{fig:2_4}
\end{figure}

In the eigenbasis of the Hamiltonian matrix~$\mat H$, the projection operator~\eqref{eq:B_proj} can be written as
\begin{equation}
    \mat P_{\varepsilon_i, \varepsilon_{i+1}} = 
    \begin{pmatrix}
        \ddots\\
        & 0 & 0 & 0 & \dots & 0 & 0 & 0 &\\
        & 0 & 1 & 0 & \dots & 0 & 0 & 0 &\\
        & 0 & 0 & 1 & \dots & 0 & 0 & 0 &\\
        & \vdots & \vdots & \vdots & \ddots & \vdots & \vdots & \vdots & \\
        & 0 & 0 & 0 & \dots & 1 & 0 & 0 &\\
        & 0 & 0 & 0 & \dots & 0 & 1 & 0 &\\
        & 0 & 0 & 0 & \dots & 0 & 0 & 0 &\\
        & & & & & & & & \ddots\\
    \end{pmatrix},
\end{equation}
i.e. its diagonals are determined by a boxcar function
\begin{equation}
    P_{\varepsilon_i, \varepsilon_{i+1}}^{jj} =
    \begin{cases}
        1, \quad \lambda_j \in [\varepsilon_i, \varepsilon_{i+1}]\\
        0, \quad \rm{otherwise}
    \end{cases},
\end{equation}
and trace of the projection matrix thus gives the number of eigenvalues inside the corresponding interval:
\begin{equation}
    \mu_{\varepsilon_i, \varepsilon_{i+1}} = \Tr(\mat P_{\varepsilon_i, \varepsilon_{i+1}}).
\end{equation}
We interpret the projection matrix as a boxcar function of the Hamiltonian matrix $\mat H$
\begin{equation}
    \mat P_{\varepsilon_i, \varepsilon_{i+1}}(\mat H)=\mathcal{H}(\mat H-\varepsilon_i\mat I)-\mathcal{H}(\mat H-\varepsilon_{i+1}\mat I)
\end{equation}
where $\mathcal{H}(x)$ is the Heaviside step function. A boxcar function can be expressed as a series of orthogonal functions. We use Chebyshev expansion due to its rapid convergence and efficient recursion relations, but in principle one could choose also other basis functions, for example, a Fourier series. We expand the projection matrix as
\begin{alignat}{3}
    \mat P_{\varepsilon_i, \varepsilon_{i+1}}(\mat H) &= \sum_{j = 0}^{\infty}\gamma_j(\varepsilon_i, \varepsilon_{i+1})\mat T_j(\mat H) \notag \\
    &\approx \sum_{j = 0}^{p}\gamma_j(\varepsilon_i, \varepsilon_{i+1})\mat T_j(\mat H),\label{eq:proj_aprox}
\end{alignat}
where~$T_j$ is the $j$-th Chebyshev polynomial of the first kind, the series is truncated to the order~$p$, and the expansion coefficients~$\gamma_j(\varepsilon_i, \varepsilon_{i+1})$ are that of the boxcar function $\gamma_0(\varepsilon_i, \varepsilon_{i+1})=[\arccos(\varepsilon_i) - \arccos(\varepsilon_{i+1})]/\pi$ and for $j\ge 1$
\begin{equation}
    \gamma_j(\varepsilon_i, \varepsilon_{i+1}) = 
        2\frac{\sin[j\arccos(\varepsilon_i)] - \sin[j\arccos(\varepsilon_{i+1})]}{j\pi}.
\end{equation}
Here we have assumed that all the eigenvalues of~$\mat H$ are inside the domain of the Chebyshev polynomials~$[-1, 1]$. We therefore first have to scale the matrix 
\begin{equation}\label{eq:2_scale}
    \mat H \to \frac{\mat H - \mat I(\lambda_{\rm max} + \lambda_{\rm min})/2}
    {\mat I(\lambda_{\rm max} - \lambda_{\rm min})/2},
\end{equation}
where~$\lambda_{\rm min}$ and~$\lambda_{\rm max}$ are the smallest and largest eigenvalues of~$\mat H$, respectively. Similar scaling has to be done also to the values~$\varepsilon_i$ and~$\varepsilon_{i+1}$. 

In principle one could obtain an estimation for the number of eigenvalues by taking the trace of the sum in Eq.~\eqref{eq:proj_aprox}. Problem with this is that in order to construct the value of the matrix-argument Chebyshev polynomial ~$\mat T_p(\mat H)$ one needs several matrix-matrix multiplications, which makes it a heavy calculation. However, we can reduce the amount of required computational resources considerably if we include the trace operation into Eq.~\eqref{eq:proj_aprox}. This replaces the matrix-matrix products with matrix-vector and vector-vector products. An option for performing the trace is to use the full computational basis set in which~$\mat H$ is expressed, but because the matrix~$\mat H$ is large, this is not very efficient. Better approach is to use a Monte--Carlo -like method. We utilize a stochastic estimator by Hutchinson~\cite{hutchinsonStochasticEstimator}, who proved that the trace of a matrix $\mat A$ can be obtained as a stochastic average of random vectors $\vec v_k$ whose elements are either~$1$ or~$-1$ with equal probabilities: $\Tr A=\lim_{M\to\infty} M^{-1} \sum_{k=1}^M \vec v_k^T\mat A \vec v_k $. For our purpose, we write the Hutchinson stochastic trace estimator as
\begin{alignat}{9}
    \Tr(\mat P_{\varepsilon_i, \varepsilon_{i+1}})
    &\approx \frac{1}{n_v}\sum_{k=1}^{n_v}\vec v_k^T\mat P_{\varepsilon_i, \varepsilon_{i+1}}\vec v_k,\label{eq:tr_approx}
\end{alignat}
where we have truncated the amount of random vectors to~$n_v$, which is much smaller than the dimension of~$\mat H$. We combine Eqs.~\eqref{eq:proj_aprox} and~\eqref{eq:tr_approx} and obtain
\begin{equation}\label{eq:ev_est}
    \mu_{\varepsilon_i, \varepsilon_{i+1}} \approx \frac{1}{n_v}\sum_{k = 1}^{n_v}\sum_{j=0}^p
    \gamma_j(\varepsilon_i, \varepsilon_{i+1})\vec v_k^T\mat T_j(\mat H)\vec v_k.
\end{equation}
Let us denote the vector~$\mat T_j(\mat H)\vec v_k$ with~$\vec w_j^k$. With the recursion relation for the Chebyshev polynomials one can write this as
\begin{equation}
    \vec w_j^k = 2\mat H\vec w_{j-1}^k - \vec w_{j-2}^k,
\end{equation}
where, since~$\mat T_0(\mat H) = \mat I$ and~$\mat T_1(\mat H) = \mat H$, we have $\vec w_0^k = \vec v_k$ and~$\vec w_1^k = \mat H\vec v_k$. Finally, Eq.~\eqref{eq:ev_est} becomes
\begin{equation}\label{eq:ev_est2}
    \mu_{\varepsilon_i, \varepsilon_{i+1}} \approx \frac{1}{n_v}\sum_{k = 1}^{n_v}\sum_{j=0}^p
    \gamma_j(\varepsilon_i, \varepsilon_{i+1})\vec v_k^T\vec w_j^k,
\end{equation}
which only contains matrix-vector and vector-vector products. In the Chebyshev expansion we have used~$n_v = 30$ random vectors and~$p = 50$ terms, which are sufficient for accurate results. Performance and accuracy comparisons between the $LDL$ decomposition and the stochastic Chebyshev expansion methods are shown in Table~\ref{tab:B_1} and Fig.~\ref{fig:2_4} showing that the Chebyshev method produces the maximum of the density of states both rapidly and accurately. 

In summary, the desired eigenstate is obtained in a following way. We first create a random realization of on-site energies~$\omega_\ell$ and construct the Hamiltonian of Eq.~\eqref{eq:2_6}. If the system is small, we use the exact $LDL$ method to construct the density of states, and otherwise we use the stochastic method of Eq.~\eqref{eq:ev_est2}. We then select the target energy at which the density of states has its maximum. This target energy is then used in the shift-and-invert algorithm to obtain the eigenpair closest to the target. The largest and smallest eigenstates, required in Eq.~\eqref{eq:2_scale}, can be obtained e.g. with power iteration, or with algorithms included in Spectra~\cite{spectraweb}. 

\begin{table}
    \centering
    \begin{tabular}{c|c|c|c}
        $L$ & $D_{L/2, L}$ & $LDL$ decomposition & Chebyshev expansion\\
        \hline
        10 & 2002 & \SI{0.02}{\second} & \SI{0.15}{\second}\\
        12 & 12376 & \SI{10}{\second} & \SI{1.1}{\second}\\
        14 & 77520 & \SI{2500}{\second} & \SI{4.7}{\second}\\
        16 & 490314 & - & \SI{41.0}{\second}\\
    \end{tabular}
    \caption{Comparison of the scaling of the execution times for estimating the number of eigenvalues within an arbitrary interval~$[\varepsilon_i, \varepsilon_{i+1}]$ based on the $LDL$ decomposition and the stochastic Chebyshev expansion. $D_{L/2, L}$ denotes the Hilbert space dimension~of~Eq.~\eqref{eq:2_states} at half-filling for different sized systems $L$. All computations are performed on the same tabletop machine using a single thread from four threaded Intel i5-2400 core with \SI{3.10}{\giga\hertz} clock frequency. The execution times of the Chebyshev expansion method scales roughly linearly with the system size due to its sparse matrix-vector multiplication.}
    \label{tab:B_1}
\end{table}

\section{Time evolution with Krylov subspace methods} \label{app:time}
The unitary time evolution of a closed quantum system is governed by the Schr{\"o}dinger equation
\begin{equation}\label{eq:4b_1}
    i\hbar\frac{d}{dt}\ket{\psi(t)} = \hop\ket{\psi(t)}.
\end{equation}
If the Hamiltonian $\hop$ is time-independent, the Schr{\"o}dinger equation has the formal solution
\begin{equation}\label{eq:4b_2}
    \ket{\psi(t)} = e^{-it\hop/\hbar}\ket{\psi_0},
\end{equation}
where $\ket{\psi_0}$ is the initial state. If one is able to diagonalize the Hamiltonian, that is, to form the eigendecomposition, the matrix exponential of Eq.~\eqref{eq:4b_2} is trivial. However, for large systems the full diagonalization is inefficient or even impossible, and therefore the matrix exponential has to be approximated. In this work we use the Krylov subspace method allowing efficient computation of the product between an exponentiated sparse matrix and a state vector~\cite{luitzErgodicSideManybody, saadIterativeMethods, molerNineteenDubiousWays2003, manmanaTimeEvolution, beerwerthKrylovSubspaceMethods2015}.

\begin{figure}
    \centering
    \includegraphics[width=1.0\linewidth]{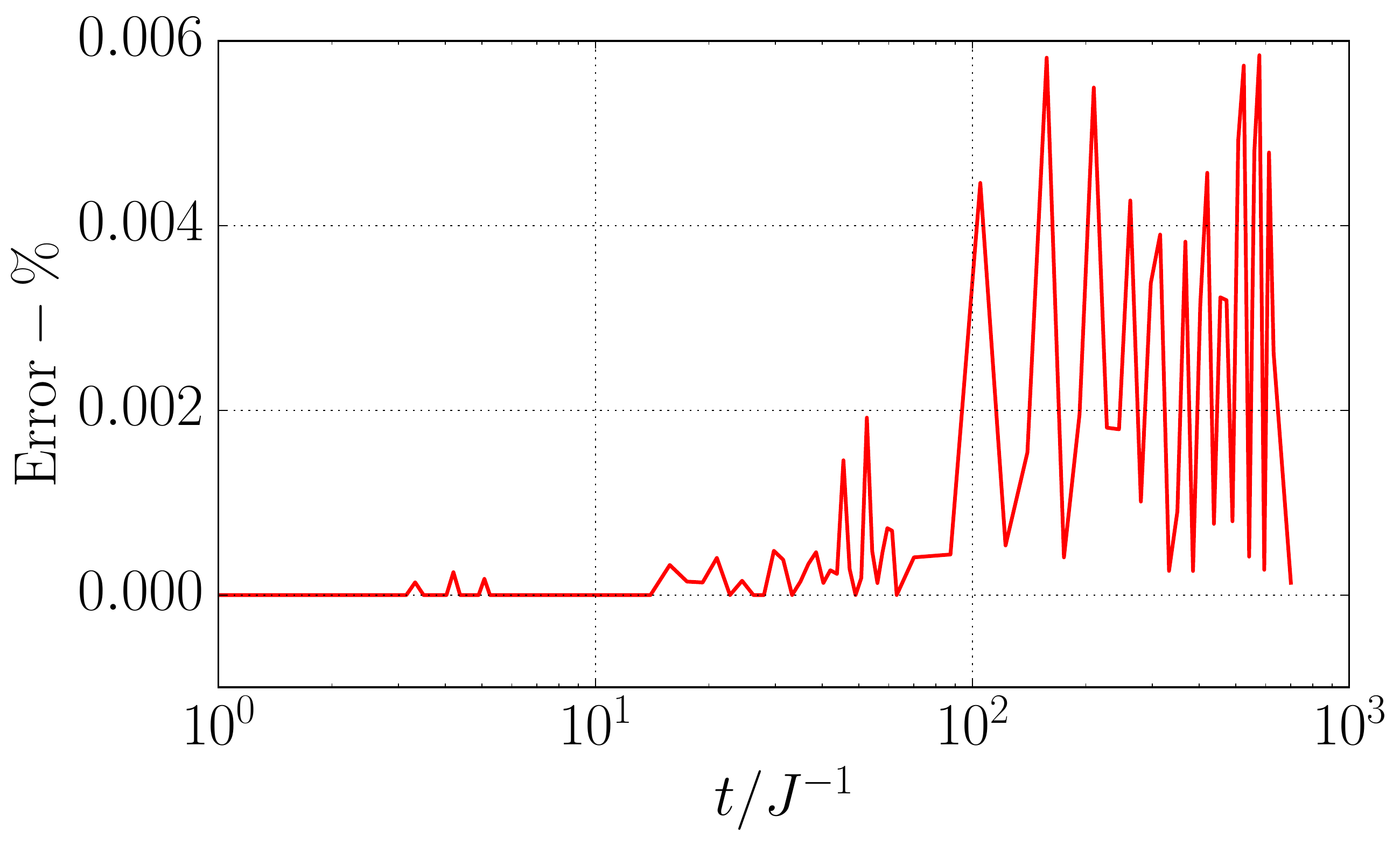}
    \caption{Relative error percent of the bipartite entanglement entropy obtained from the exact diagonalization and Krylov method with~$m=5$  as a function of evolution time calculated for a single, representative disorder realization. The system parameters of Eq.~\eqref{eq:2_6} are~$L=12$,~$W/J = 15$, and~$U/J = 3.5$.}
    \label{fig:4b_1}
\end{figure}

If the system at time~$t$ is in state~$\ket{\psi(t)}$, after a time~$\tau$ the state becomes
\begin{equation}\label{eq:4b_3}
    \ket{\psi(t+\tau)} = e^{-i\tau\hop/\hbar}\ket{\psi(t)},
\end{equation}
If the time step~$\tau$ is short, one can accurately express the state vector~$\ket{\psi(t)}$ and the Hamiltonian~$\hop$ in an $m$-dimensional Krylov subspace $\mathcal{K}_{m}$, with $m$ much smaller than the Hilbert space dimension. This subspace is spanned by the vectors
\begin{displaymath}
    \left\{\vec v_0, \mat H\vec v_0, \mat H^2\vec v_0,\dots,\mat H^{m-1}\vec v_0\right\},
\end{displaymath}
where the vector~$\vec v_0$ denotes the state~$\ket{\psi(t)}$ and $\mat H$ is the Hamiltonian in the matrix form. Because $\mat H$ is Hermitian, an orthogonal matrix 
\begin{equation}
     \mat K_m = 
     \begin{pmatrix}
        \vec v_0 & \vec v_1 & \vec v_2 & \dots & \vec v_{m-1}
     \end{pmatrix}
\end{equation}
can be constructed with the Lanczos-algorithm~\cite{luitzErgodicSideManybody, saadIterativeMethods, molerNineteenDubiousWays2003, manmanaTimeEvolution, beerwerthKrylovSubspaceMethods2015, lanczos50}. The Hamiltonian in the Krylov subspace then becomes a tridiagonal~$m\times m$ matrix
\begin{equation}
    \mat K_m^\dagger\mat H\mat K_m = \mat M_m = 
    \begin{pmatrix}
        \alpha_0 & \beta_1 & & & 0\\
        \beta_1 & \alpha_1 & \beta_2 & \\
        & \beta_2 & \alpha_2 & \ddots\\
        &  & \ddots & \ddots & \beta_{m-1} \\
        0& & & \beta_{m-1} &\alpha_{m-1}\\
    \end{pmatrix}.
\end{equation}
The matrix elements~$\alpha_j$ and~$\beta_j$ as well as the orthogonal vectors~$\vec v_j$ are obtained from equations~\cite{saadIterativeMethods, beerwerthKrylovSubspaceMethods2015}
\begin{subequations}
\begin{alignat}{9}
    \alpha_j &= \vec v_j\cdot\left(\mat H\vec v_j - \beta_{j-1}\vec v_{j-1}\right),\\
    \beta_j\vec v_{j+1} &= \mat H\vec v_j - \alpha_j\vec v_j - \beta_{j-1}\vec v_{j-1}.
\end{alignat}
\end{subequations}
After solving the matrix $\mat K_m$, the approximative time evolution can be calculated in the Krylov subspace as~\cite{luitzErgodicSideManybody, molerNineteenDubiousWays2003}
\begin{equation}
    \ket{\psi(t+\tau)} = e^{-i\tau\hop/\hbar}\ket{\psi(t)} \approx 
    \mat K_me^{-i\tau\mat M_m/\hbar}\mat K_m^\dagger\vec v_0,
\end{equation}
where the exponential of the small tridiagonal matrix~$\mat M_m$ is easily computed either with the eigendecomposition or Pad\'e approximation provided by many numerical libraries. The Krylov method gives accurate results because the eigenvalues of the tridiagonal matrix approximate the eigenvalues of the Hamiltonian that are the most important for the dynamics during the current time step~\cite{molerNineteenDubiousWays2003}. 

The Lanczos algorithm used for constructing the basis of the Krylov subspace can suffer from numerical problems. The floating-point arithmetic causes numerical instabilities which lead to a loss of orthogonality of the vectors~$\vec v_j$ with increasing subspace dimension~$m$~\cite{saadIterativeMethods, beerwerthKrylovSubspaceMethods2015}. This can be fixed by performing a re-orthogonalization on the matrix~$\mat K$. However, for our simulations the Krylov method produces sufficiently accurate results before the orthogonality is lost. Error of the Krylov method as compared to the exact diagonalization is shown in Fig.~\ref{fig:4b_1}. We see that the Krylov method is very accurate even for very small subspace size~$m=5$. The error is cumulative, which restricts this method to shorter times than the exact diagonalization.

\section{Time evolving block decimation with matrix product states} \label{app:timeMPS}

\begin{figure}
    \centering
    \includegraphics[width=1.0\linewidth]{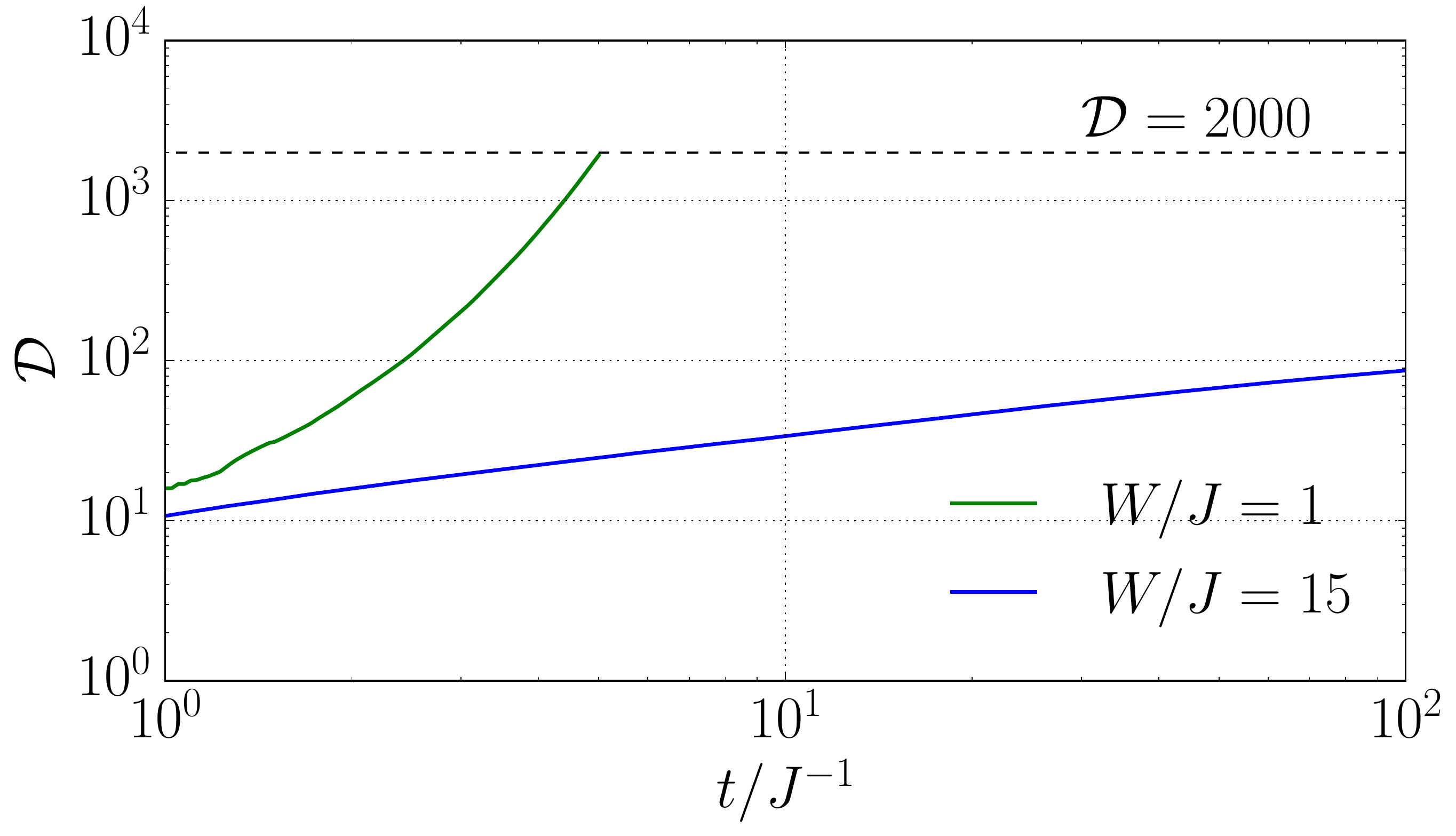}
    \caption{Evolution of the average bond dimension~$\mathcal{D}$ in ergodic (green) and many-body localized (blue) phases.}
    \label{fig:bond}
\end{figure}

For long chains of transmons, we express the wave function $\ket{\psi}$ in terms of matrix product states (MPS) and then utilize an approximate method, denoted as the time evolving block decimation, to efficiently calculate the unitary time-evolution of Eq.~\eqref{eq:4b_3}. First, matrix product states for an open chain of $L$ transmons are defined as, 
\begin{equation}\label{Eq:MPS}
\ket{\psi}=\sum_{\{n\}}A^{n_{1}}_{1}A^{n_{2}}_{2}\ldots A^{n_{L}}_{L}
  \ket{n_1 n_2\ldots n_{L}},
\end{equation}
where the index $n_{\ell}\in 0,1,\ldots ,n_{\rm max}$ labels the number of bosons at site $\ell$ with maximum (truncated) occupation~$n_{\rm max}$ and the sum is performed over all ``physical" indices $n_{\ell}$. $A^{n_{\ell}}_{\ell}$ is a $N_{\ell} \times M_{\ell}$ matrix. For the left/right boundary tensors, $N_{1}$, $M_{L}=1$. The dimensions associated to $N_{\ell}$ and $M_{\ell}$ are the ``bond" dimensions between sites $\ell-1,\ell$ and $\ell,\ell+1$, respectively. In numerical algorithms, we bound the size of the matrix  $N_{\ell},M_{\ell}<\mathcal{D}$. The bound on bond dimension, for example $N_{\ell}$, is equivalent to a bound on entanglement entropy $S \sim \ln{\mathcal{D}}$, across the bipartition $(1,\ell-1)\cup (\ell,L)$.

The total number of resources $\mathcal{R}$ required to store a matrix product state is $\mathcal{R} \sim n_{\rm max} \mathcal{D}^2 L$. If the entanglement entropy is small,  matrix product states provide an efficient classical storage container of many-body quantum states. Examples of such states are gapped ground states, many-body localized eigenstates and in general any state which obeys area-law in entanglement entropy, i.e.~$S \sim \text{const}$ for all bipartitions $(1,\ell-1)\cup(\ell,L)$. Unfortunately, entanglement entropy for a quench at finite energy density in a chaotic system increases linearly with time~$S\propto t$, which implies that for a fixed cutoff $\mathcal{D} \sim \mathcal{O}(10^3)$ the state can be quantitatively approximated up to times for which $T\propto\ln(\mathcal{D})$. On the other hand, for many-body localized systems~$S\propto\ln{t}$, which implies $T\propto \mathcal{D}^c$, where $c$ is a disorder dependent constant. In Fig.~\ref{fig:bond} we compare the computational resources over time in the ergodic and many-body localized phases by monitoring the growth of average bond dimension over time. We track the largest bond dimension per disorder realization and then average over realizations. As expected, the scaling in the localized phase is a power-law $\mathcal{D}_{\rm{loc}}\propto t^a$, while the scaling in the ergodic phase is exponential $\mathcal{D}_{\rm{erg}}\propto e^{a t}$. 

Time evolving block decimation~\cite{vidal2004efficient} is an evolution scheme based on the application of Trotter formula to the unitary time evolution, i.e. breaking the many-body unitary to a successive application of few-body unitary gates, and a controlled truncation of the matrix product state after the application of each gate. 

\subsection{Integrator}
We first describe the integrator we use, which  is a fourth order trotterization scheme. For a nearest-neighbor Hamiltonian we define the ``forward/backward'' sweeps (time-evolution steps) as,
\begin{subequations}
\begin{align}
&\Phi(dt) = U_{L,L-1}(dt)U_{L-1,L-2}(dt),\ldots, U_{2,1}(dt),\\
&\Phi^*(dt) = U_{1,2}(dt)U_{2,3}(dt),\ldots, U_{L-1,L}(dt),
\end{align}
\end{subequations}
where the two-site gates are defined as
\begin{equation}
U_{\ell,\ell+1}(dt) = e^{-i h_{\ell,\ell+1}dt}.
\end{equation}
The Hamiltonian density $h_{\ell,\ell+1}$, is defined to symmetrically include the one-site terms of the Hamiltonian of Eq.~(\ref{eq:2_6}),
\begin{subequations}
\begin{align}
     h_{\ell,\ell+1} &= \frac{1}{2}(h_{\ell}+h_{\ell+1}) + J_\ell\left(\aop_{\ell}^\dagger\aop^{}_{\ell+1} +h.c.,\right),\\
     h_{\ell} &= \omega_\ell\nop_\ell-\frac{U_\ell}{2}\nop_\ell\left(\nop_\ell - 1\right).
\end{align}
\end{subequations}
The backward sweep is the adjoint method of the forward sweep, i.e. $\Phi^*(-dt)\Phi(dt)\ket{\psi}=\ket{\psi}$. It can be shown that for self-adjoint methods the Trotter error is always of even order~\cite{marsden1993interdisciplinary}. Thus, it is favorable to create the self-adjoint method $\Psi(dt)=\Phi^*(dt/2)\Phi(dt/2)$, which is a second order method. Using standard composition rules~\cite{marsden1993interdisciplinary}, we compose the following fourth order self-adjoint method,
\begin{equation}
 F(dt) =   \Psi(a_{1} dt)\Psi(a_{2} dt)\Psi(a_{1} dt),
\end{equation}
where, $a_{1}= (2-2^{1/3})^{-1}$, $a_2 = 1- 2 a_{1}$ and ${F(T) = e^{-i H  T/\hbar}+ T \mathcal{O}(dt^4)}$.

\subsection{Truncation}
Matrix product states have some gauge freedom associated to the virtual bond dimension. We use the ``mixed" gauge~\cite{SCHOLLWOCK201196} to perform the evolution and efficiently truncate the matrix product state. We discuss the process for a forward sweep. Before applying the gate $U_{\ell,\ell+1}$ mixed gauge implies that local tensors satisfy
\begin{subequations}
\begin{align}
 &\sum_{n_{k}}A^{n_{k}\dag}_{k}A^{n_{k}}_{k} = 1\!\!1_{M_{k}\times M_{k}}, \; k < \ell,\\
 &\sum_{n_{k}}A^{n_{k}}_{k}A^{n_{k}\dag}_{k} = 1\!\!1_{N_{k}\times N_{k}}, \; k > \ell,
\end{align}
\end{subequations}
while the tensor at site $\ell$ does not have any special property. The tensors at site $\ell,\ell+1$ are combined and the two-body gate is applied,
\begin{equation}
\Theta_{\ell,\ell+1} = U_{\ell,\ell+1}\sum_{n_{\ell},n_{\ell+1}} A^{n_{\ell}}_{\ell}A^{n_{\ell+1}}_{\ell+1}\ket{n_{\ell} n_{\ell+1}}.
\end{equation}
The updated local tensors are constructed from the singular value decomposition,
\begin{subequations}\label{eq:svd}
\begin{align}
\Theta_{(n_{\ell}b_{\ell}),(n_{\ell+1}c_{\ell+1})}  &= \sum_{a}\mathcal{U}_{(n_{\ell}b_{\ell}),a}S_{a,a}\mathcal{V}^{*}_{a,(n_{\ell+1}c_{\ell+1})},\\
A^{n_{\ell}}_{b_{\ell},a_{\ell}} &=\mathcal{U}_{(n_{\ell}b_{\ell}),a},\\
A^{n_{\ell+1}}_{a', c_{\ell+1}} &= \sum_{a}S_{a',a}  \mathcal{V}^{*}_{a,(n_{\ell+1}c_{\ell+1})},
\end{align}
\end{subequations}
Due to the unitarity of the matrix $\mathcal{U}$, $\mathcal{U}^{\dag}\mathcal{U} = 1\!\!1$, the updated tensor $A_{\ell}$ obeys the correct  gauge form required for the evolution of sites of $\ell+1,\ell+2$. The singular values $\lambda_{k}$ (in descending order) are the diagonal elements of $S$ and correspond to the square roots of the eigenvalues of the reduced density matrix $\rho = \Tr_{1\ldots \ell}\ket{\psi}\bra{\psi}$, thus $\sum_{k}\lambda^2_{k} = 1 $. We truncate the singular values for $k>k_{c}$ so that the probability loss  $\sum_{k>k_{c}}\lambda^2_{k}< \epsilon$, and then renormalize the truncated density matrix, $\sum_{k\leq k_{c}}\lambda^2_{k} = 1$. 
This process is iterated to perform a forward sweep. The backward sweep is performed in a similar way. The computational bottleneck of the algorithm is the singular value decomposition of Eq.~\eqref{eq:svd} which scales as $\mathcal{O}[(n_{\rm max}\mathcal{D})^3]$ for a square matrix of dimension $n_{\rm max} \mathcal{D}$.

\subsection{Numerical details}\label{sub:numerical}
Calculations were performed using the ITensor Library~\cite{ITensor}. In all simulations we have explicitly conserved the $U(1)$ symmetry associated to the total particle number, which allows for an additional speedup due to the block diagonal structure of the matrix $\Theta$. For the simulations in the ergodic regime ($W/J=1$, $U/J=3.5$) we have used time step $dt/J^{-1}= 0.025 $ and cutoff $\epsilon = 10^{-8}$ as well as a hard cutoff to the bond dimension $\mathcal{D}_{c} = 2000$. The local Hilbert space is truncated at $n_{\rm max} = 4$. The simulations are stopped at time $T/J^{-1} = 5 $, where the hard cutoff is reached for most realizations to ensure that the quantities of interest are accurate. 
For the disordered regime ($W/J=15$, $U/J=3.5$) we have used $dt/J^{-1}= 0.01$ and cutoff $\epsilon = 10^{-9}$ as well as a hard cutoff to the bond dimension $\mathcal{D}_{c} = 2000$. To gain additional speedup we truncated the local Hilbert space at $n_{\rm max}= 3$ which does not affect the dynamics for strong disorder and dilute initial state. The simulations are stopped at time $T/J^{-1}=200$. The bond dimension is not saturated for any disorder realization.


%

\end{document}